\documentclass[twocolumn]{aastex631}

\usepackage{graphicx}
\usepackage{amsmath,amssymb}
\usepackage[T1]{fontenc}
\usepackage{longtable}

\begin{document}

\title{Dense Cores in IRDC G14.225-0.506 revealed by ALMA Observations}

\author[0009-0009-1522-2240]{Yanhanle Zhao}
\affiliation{Department of Physics, The Chinese University of Hong Kong, Shatin, New Territories, Hong Kong SAR} 
\affiliation{Center for Astrophysics $|$ Harvard $\&$ Smithsonian, 60 Garden Street, Cambridge, MA 02138, USA}

\author[0000-0003-2384-6589]{Qizhou Zhang}
\affiliation{Center for Astrophysics $|$ Harvard $\&$ Smithsonian, 60 Garden Street, Cambridge, MA 02138, USA}

\author[0000-0002-4774-2998]{Junhao Liu}
\affiliation{National Astronomical Observatory of Japan, 2-21-1 Osawa, Mitaka, Tokyo 181-8588, Japan}

\author[0000-0003-1337-9059]{Xing Pan}
\affiliation{Center for Astrophysics $|$ Harvard $\&$ Smithsonian, 60 Garden Street, Cambridge, MA 02138, USA}
\affiliation{School of Astronomy and Space Science, Nanjing University, 163 Xianlin Avenue, Nanjing 210023, P.R.China}
\affiliation{Key Laboratory of Modern Astronomy and Astrophysics (Nanjing University), Ministry of Education, Nanjing 210023, P.R.China}

\author[0000-0001-6924-9072]{Lingzhen Zeng}
\affiliation{Center for Astrophysics $|$ Harvard $\&$ Smithsonian, 60 Garden Street, Cambridge, MA 02138, USA}

\begin{abstract}

Dense cores in massive, parsec-scale molecular clumps are sites that harbor protocluster formation. We present results from observations towards a hub-filament structure of a massive Infrared Dark Cloud (IRDC) G14.225-0.506 using the Atacama Large Millimeter/submillimeter Array (ALMA). The dense cores are revealed by the 1.3 mm dust continuum emission at an angular resolution of $\sim$ 1.5$''$ and are identified through the hierarchical Dendrogram technique. Combining with the N$_2$D$^+$ 3-2 spectral line emission and gas temperatures derived from a previous NH$_3$ study, we analyze the thermodynamic properties of the dense cores. The results show transonic and supersonic-dominated turbulent motions. There is an inverse correlation between the virial parameter and the column density, which implies that denser regions may undergo stronger gravitational collapse. Molecular outflows are identified in the CO 2-1 and SiO 5-4 emission, indicating active protostellar activities in some cores. Besides these star formation signatures revealed by molecular outflows in the dense cores, previous studies in the infrared, X-ray, and radio wavelengths also found a rich and wide-spread population of young stellar objects (YSOs), showing active star formation both inside and outside of the dense cloud.

\end{abstract}

\section{Introduction} \label{sec:intro}

Infrared dark clouds (IRDC) are typically made of cold ($<$25 K) and dense ($>$10$^4$ cm$^{-3}$) molecular gas seen against the Galactic background radiation in the infrared wavelength. They are widely used in investigating the early evolutionary stages of star formation \citep{Carey_2000, Rathborne_2007, Peretto_2013, Barnes_2021}. IRDCs with masses larger than 10$^3$ M$_\odot$ and sizes of $\sim$1 pc are referred as clumps \citep{Zhang_2009} and are candidates for studies of massive star ($>$ 10 M$_\odot$) and protocluster formation \citep{Zhang_2015, Traficante_2023}. The properties of IRDC clumps include lower temperatures of $\sim$15 K, lower luminosities, a limited detection rate of H$_2$O masers \citep{Wang_2006} and narrower linewidths of the order of 2 km~s$^{-1}$ compared to the high-mass protostellar objects (HMPOs) and H II regions. The latter objects exhibit active massive star formation with a luminosity $>$10$^4$ L$_\odot$, and contains emission from complex organic molecules \citep{van_Dishoeck_1998} and energetic outflows \citep{Rathborne_2011, Zhang_2001, Beuther_2002}. The difference in physical and chemical properties implies that HMPOs and H II regions are more evolved than the IRDC clumps. Therefore, these massive IRDC clumps can be primary targets for investigating the very early evolution of massive star formation \citep{Rathborne_2007}.

Sensitive and spatially resolved observations of clumps reveal smaller-scale and higher-density entities -- dense cores with typical sizes of $\sim$0.01 - 0.1 pc where individual or a small group of stars form. Studies of these cores can provide perspectives toward understanding their physical and chemical states and properties of the initial stages in massive star formation \citep{Williams_2018, Barnes_2021}. Under extreme conditions of low temperatures and high densities \citep{Caselli_2002}, substantial CO depletion and deuteration fractionation are expected \citep{Pillai_2012, Chen_2010}. Thermodynamic analysis of some dense cores reveals their sub-virial properties with a virial mass smaller than the gas mass of the cores~\citep{Lu_2014, Li_2013, Zhang_2015, Ohashi_2016}. However, other factors such as the external pressure and the magnetic field could alter the virial state of dense cores, since external pressure acts on cores embedded in the clumps and thus contributes to the imbalance between the gravitational pull and the thermal support, while the magnetic field provides additional support for the core structures \citep{Barnes_2021}, making the conditions of virial balance in these cores more complex. 

This study analyzes the IRDC G14.225-0.506 (hereafter G14.225) as part of the M17 SWex IRDC complex located at a distance of 1.98$^{+0.13}_{-0.12}$ pc \citep{Xu_2011}. Among the known IRDCs in the Galaxy, G14.225 stands out for the presence of a network of parallel filaments as revealed by the NH$_3$ (1, 1) emission \citep{Busquet_2013}. The plane-of-the-sky component of the magnetic field is found to be mostly uniformly distributed in an orientation perpendicular to the long axis of the parsec-scale filaments \citep{Santos_2016, Anez-Lopez_2020}, which indicates that the magnetic field plays a dynamically important role in the formation of the parallel filaments \citep{Van_Loo_2014}. The NH$_3$ filaments appear to be connected with two warm hubs ("hub-filament systems"), and the velocity fields revealed in the N$_2$H$^+$ 1-0 emission indicate mass flows toward the hubs where cluster star formation takes place \citep{Chen_2019}. 
% and exhibits a network of parallel filaments with two warm hubs ("hub-filament systems") as revealed by the NH$_3$ (1, 1) emission \citep{Busquet_2013}. 
A dynamical analysis shows a $\sim$2-times larger velocity dispersion in the hubs than along the filaments and a general virial equilibrium measured from the N$_2$H$^+$ filaments \citep{Chen_2019}. \cite{Ohashi_2016} have investigated the dense core properties in IRDC G14.225 using the 3 mm continuum emission at an angular resolution of $\sim$3$''$ obtained with the Atacama Large Millimeter/submillimeter Array (ALMA). They identified 20 protostellar cores out of 48 core candidates in the 3 mm continuum emission. An analysis of virial parameters shows a dominant sub-virial trend that indicates possible collapse in these cores. Surveys of young stellar objects (YSOs) have been conducted in the infrared, X-ray, and radio wavelengths toward the G14.225 region. A YSO survey was conducted with the $Sptizer$ GLIMPSE and MIPSGAL data by \cite{Povich_2010} and the IR catalog was supplemented by the $Chandra$ X-ray and UKIDSS Galactic Plane Survey \citep{Povich_2016}. Recently, \cite{Elena_2024} analyzed properties of radio sources and the signatures of star-forming activities using the continuum survey at 6 and 3.6 cm with the VLA. They found a steeper YSO mass function (YMF) as compared to the initial mass function (IMF), which implies a deficit of high-mass YSOs \citep{Ohashi_2016}. The widely-detected X-ray emission in the region reveals a population of intermediate-mass pre-main-sequence stars (\cite{Povich_2016}). Studies by \citet{Povich_2010, Povich_2016, Ohashi_2016, Chen_2019} suggest that there is an absence of massive O-type stars or massive protostars and massive cores in the G14.225 region. 

In this paper, we report sensitive observations with ALMA in the 1.3 mm continuum emission and spectral line emissions in ALMA band 6 (230 GHz) at an angular resolution of $\sim$1.5$''$. Molecular outflows are identified for the first time in this region using the CO 2-1 and SiO 5-4 lines. The dense-gas tracer N$_2$D$^+$ 3-2 is also applied in the thermodynamic analysis. The paper is structured as follows: Section \ref{sec:obs} introduces the observations; Section \ref{sec:results} presents the analysis of the dust continuum and line emission data; Section \ref{sec:discussion} discusses possible implications of the observational results, and Section \ref{sec:conclusion} summarizes the main findings.

\begin{figure}[ht!]
\includegraphics[width=8cm]{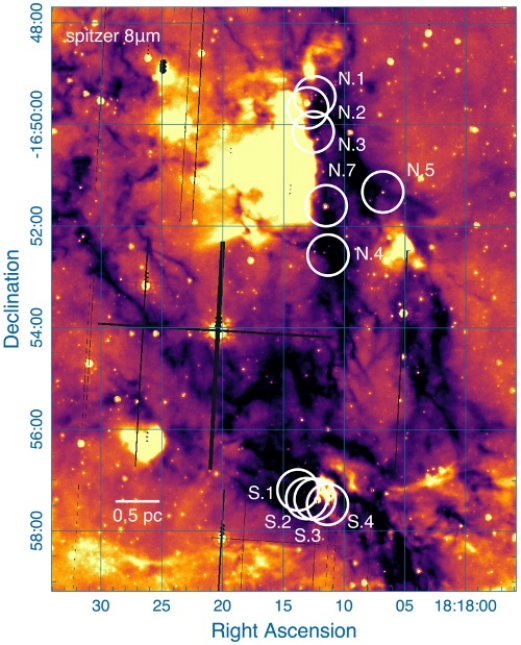}
\caption{Observed fields (white circles corresponding to the FWHM of the primary beam) overlaid on the 8 $\mu$m Spitzer map (color scale). N mosaic field combines G14.225N.1, G14.225N.2, and G14.225N.3 field, and single pointing fields are G14.225N.4, G14.225N.5, and G14.225N.7. S mosaic field consists of G14.225S.1, G14.225S.2, G14.225S.3 and G14.225S.4 field.}
\label{fig:pointing}
\end{figure}

\section{Observations} \label{sec:obs}

IRDC G14.225 was observed with ALMA on July 17, 2018 under project 2017.1.00793.S (Cycle 5; PI: Qizhou Zhang). A total of 47 12m diameter antennas were employed during the observations that lasted 3 hours and 27 min. 7m ACA array and TP array are not included in this observation. As shown in Figure \ref{fig:pointing}, observations covered 10 pointings: 6 northern fields including 3 pointings (G14.225N.1, N.2, N.3) combined into the N mosaic field, 3 single-pointing for the N.4, N.5 and N.7 fields, and a 4-pointing mosaic (G14.225S.1, S.2, S.3, S.4) in the south. The positions of the corresponding pointing centers are summarized in Table \ref{tab:pointing}. Band 6 receivers were employed for the project. Three basebands, each with a bandwidth of 1.875 GHz, cover frequencies $\sim$215.5-217.4, $\sim$217.6-219.5 GHz and $\sim$232.6$-$234.5 GHz. Additionally, 4 narrow spectral windows centered at frequencies of 230.5 GHz, 231.0 GHz, 231.2 GHz, and 231.3 GHz are adopted to target the spectral line emission from the CO 2-1, OCS 19-18, $^{13}$CS 5-4 and N$_2$D$^+$ 3-2 transition, respectively. These spectral windows have a channel width of 122 kHz and a bandwidth of 58.6 MHz. The SiO J=5-4 transition at a rest frequency of 217.10 GHz is covered in one of the continuum spectral windows with a channel spacing of 0.98 MHz. The systematic velocity of all the fields is 20 km~s$^{-1}$.

Data reduction and calibration were performed using the Common Astronomy Software Applications (CASA). The continuum and spectral line emissions were separated by subtracting the continuum in line-free channels from the visibility data. Maps of the continuum and spectral line emission were generated by $TCLEAN$ using briggs weighting with a robust parameter of 0.5, and a pixel size of $0.''2$. The synthesized beam sizes of each field range from $1.''45\times 1.''00$ to $1.''47\times 1.''02$ for the continuum, and from $1.''42\times 0.''98$ to $1.''46\times 1.''03$ for the CO line emission, respectively. The root mean square (rms) noise level before the primary beam correction is $\sim$0.05-0.10 mJy~beam$^{-1}$ for the continuum emission and $\sim$2-4 mJy~beam$^{-1}$ for the CO line emission at a resolution of 0.3 km~s$^{-1}$. The spectral setup of the observations and imaging parameters are listed in Table~\ref{tab:setup}.

\begin{table}[h!]
\centering
\caption{Position of the fields}
\begin{tabular}{|c|c|c|}
\hline
Field Name & R.A.        & Decl.        \\ 
           & (h:m:s)     & (d:m:s)      \\ \hline \hline
G14.225S.1 & 18:18:13.80 & -16:57:11.50 \\ \hline
G14.225S.2 & 18:18:13.08 & -16:57:21.60 \\ \hline
G14.225S.3 & 18:18:12.39 & -16:57:22.40 \\ \hline
G14.225S.4 & 18:18:11.36 & -16:57:28.50 \\ \hline
G14.225N.1 & 18:18:12.44 & -16:49:27.10 \\ \hline
G14.225N.2 & 18:18:13.02 & -16:49:40.20 \\ \hline
G14.225N.3 & 18:18:12.57 & -16:50:08.60 \\ \hline
G14.225N.4 & 18:18:11.34 & -16:52:34.00 \\ \hline
G14.225N.5 & 18:18:06.84 & -16:51:19.40 \\ \hline
G14.225N.7 & 18:18:11.50 & -16:51:35.70 \\ \hline
\end{tabular}
\label{tab:pointing}
\end{table}

\begin{table*}[]
\centering
\caption{Observational Setup and Parameters}
\begin{tabular}{|c|c|c|c|}
\hline
Emission       & Center frequency & Bandwidth & Channel spacing \\
& (GHz) & (MHz) & (MHz) \\
\hline
continuum & 216.44  & 1875.0 & 0.98  \\
continuum  & 218.54 & 1875.0  & 0.98  \\
continuum & 233.55 & 1875.0 & 0.98 \\
\hline
CO 2-1  & 230.54 & 58.6  & 0.12  \\
OCS 19-18 & 231.06 & 58.6 & 0.12 \\
$^{13}$CS 5-4  & 231.22 & 58.6 & 0.12 \\
N$_2$D$^+$ 3-2 & 231.32 & 58.6  & 0.12  \\
\hline
\end{tabular}

\begin{tabular}{|c|c|c|c|c|c|}
\hline
 & Synthesized beam size & rms & Synthesized beam size & Chanel width & rms  \\
Field & Continuum & Continuum & CO 2-1 & CO 2-1 & CO 2-1\\
& ( $'' \times \ ''$)& (mJy beam$^{-1}$) & ( $'' \times \ ''$) & (km s$^{-1}$) & (mJy beam$^{-1}$)\\
\hline
N.4 & 1.$''$45$\times$1.$''$00 & $\sim$0.10 & 1.$''$42$\times$0.$''$98 & 0.30 & $\sim$3.5\\ 
N.5 & 1.$''$46$\times$1.$''$00 & $\sim$0.05 & 1.$''$43$\times$0.$''$98 & 0.30 & $\sim$3.5 \\
N.7 & 1.$''$46$\times$1.$''$00 & $\sim$0.05 & 1.$''$43$\times$0.$''$98 & 0.30 & $\sim$3.5\\
Mosaic N (N.1, N.2, N.3) & 1.$''$47$\times$1.$''$02 & $\sim$0.10 & 1.$''$46$\times$1.$''$03 & 0.30 & $\sim$3.0\\
Mosaic S (S.1, S.2, S.3, S.4) & 1.$''$46$\times$1.$''$01 & $\sim$0.06 & 1.$''$46$\times$1.$''$02 & 0.30 & $\sim$2.5\\
\hline
\end{tabular}
\label{tab:setup}
\end{table*}

\section{Results} \label{sec:results}

\subsection{Identification of Dense Core Structures}

The {\fontfamily{qcr}\selectfont Astrodendro}\footnote{\url{http://dendrograms.org/}} package \citep{astrodendro_2019} was applied to identify dense core structures from the 1.3 mm continuum emission. This algorithm identifies hierarchical structures including trunks (which have no parent structures), branches (stemming from the trunk), and leaves (with no sub-structure) based on the structure of intensities. Parameters used for this identification are the minimum level included in the dendrogram {\fontfamily{qcr}\selectfont min\_value}$=3\sigma$; significance in difference between the peak intensity and the level merged into its parent structure or the step size {\fontfamily{qcr}\selectfont min\_delta}$=\sigma$, where $\sigma$ is the rms of the noise level in the continuum map before the primary beam correction; and the minimum number of pixels required for a {\fontfamily{qcr}\selectfont leaf} to be considered independent as an entity, which is approximately the area of the synthesized beam size {\fontfamily{qcr}\selectfont min\_npix}$=28$. 

\begin{figure*}[ht!]
\includegraphics[width=5.95cm]{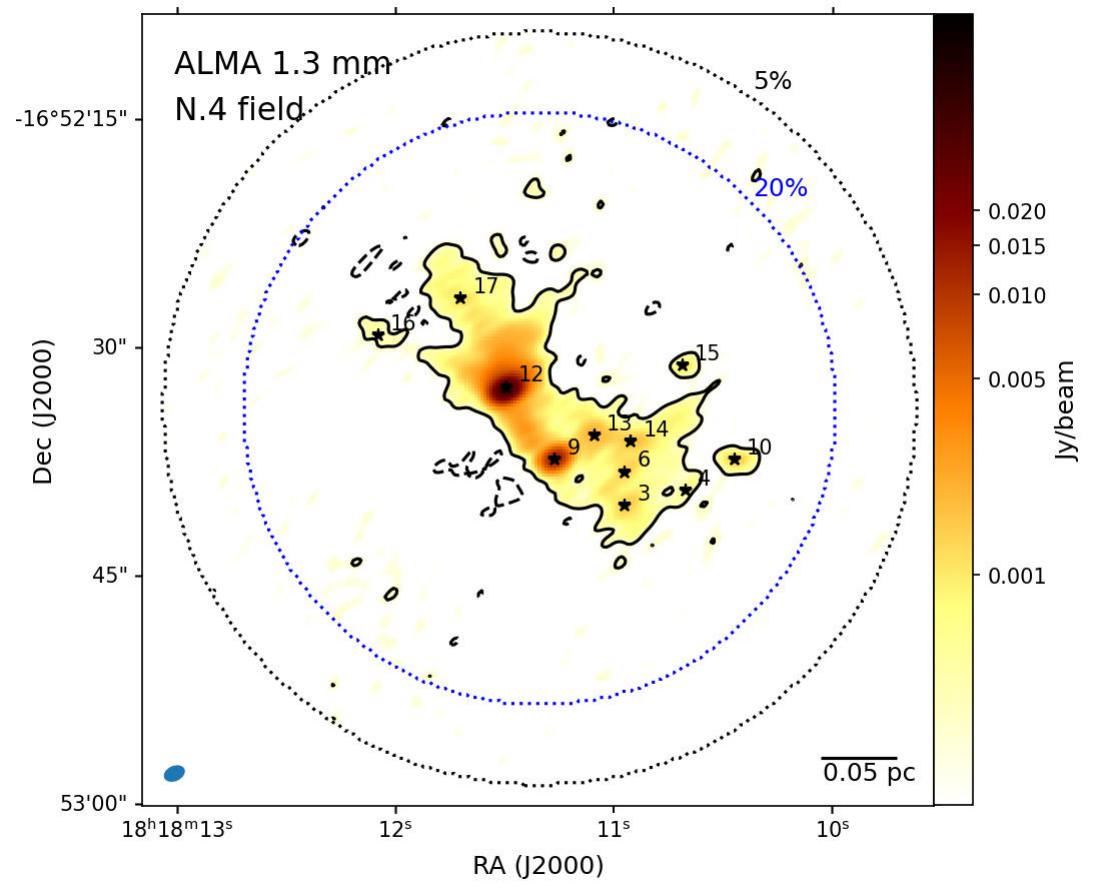}
\includegraphics[width=5.99cm]{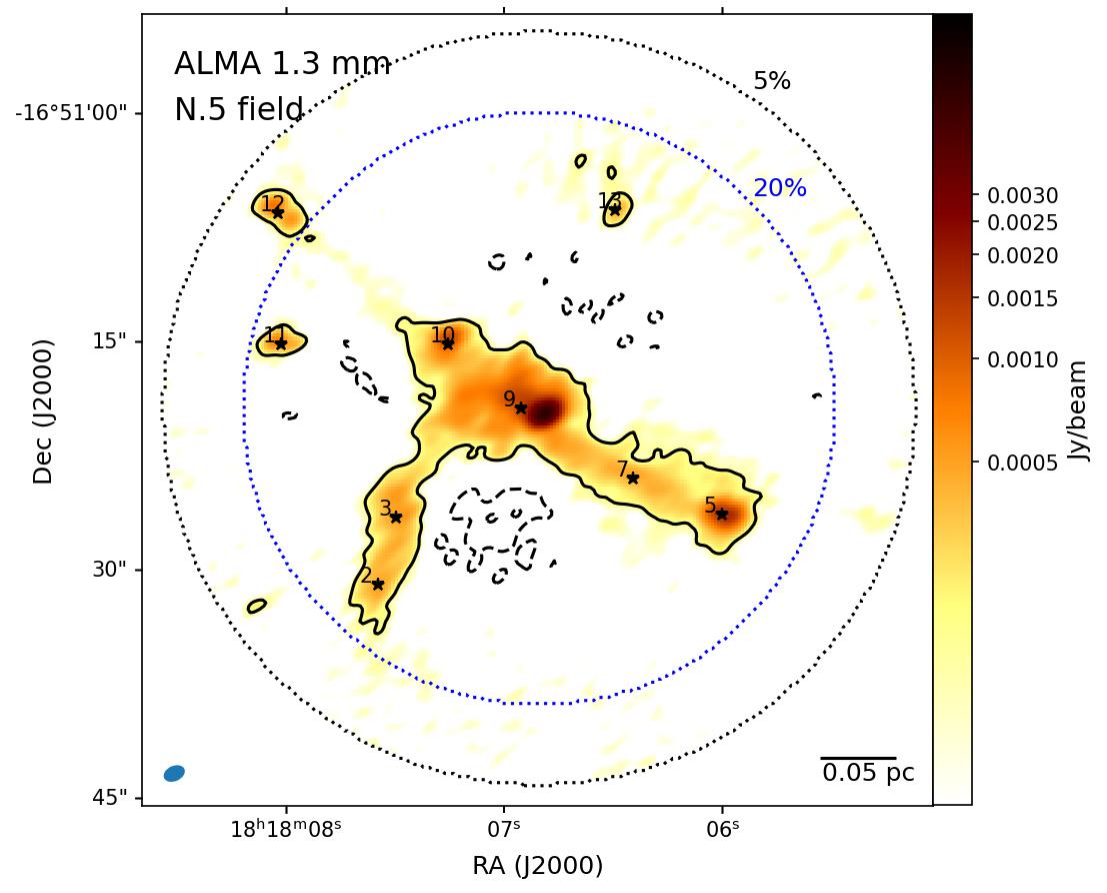}
\includegraphics[width=5.95cm]{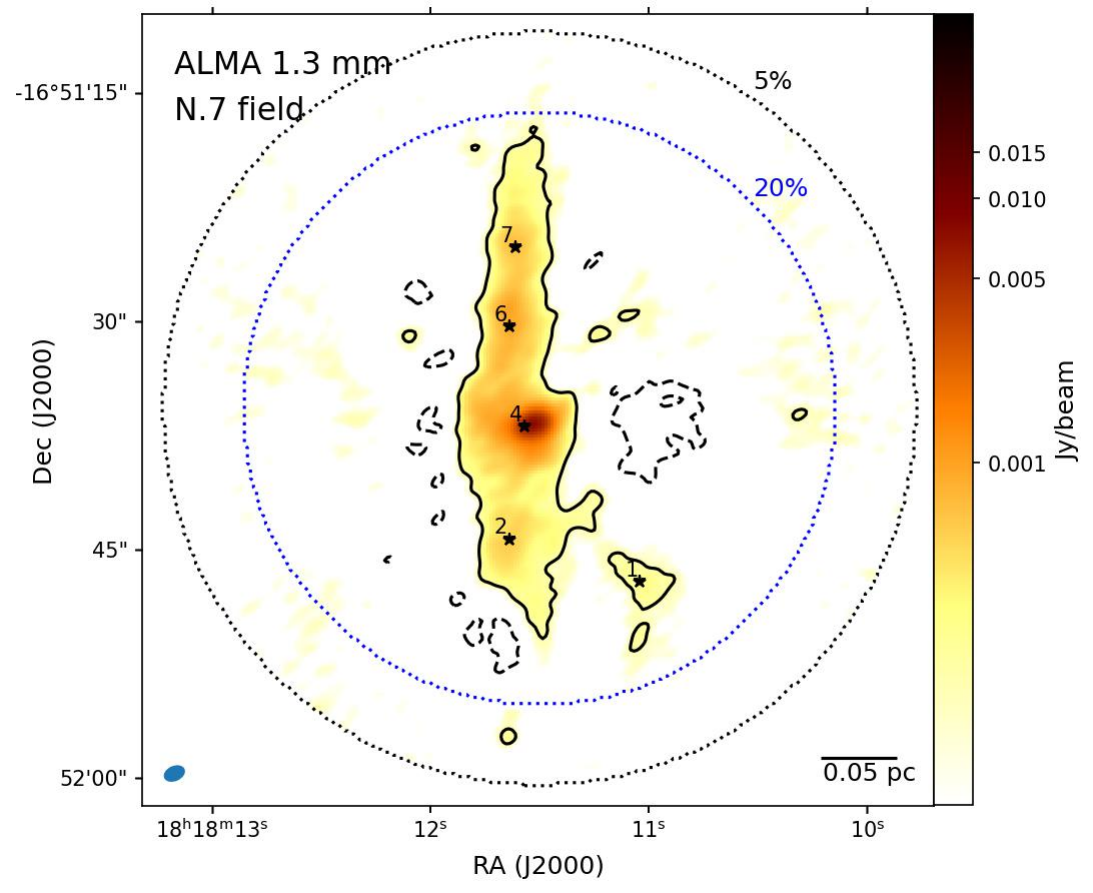}
\includegraphics[width=7.9cm]{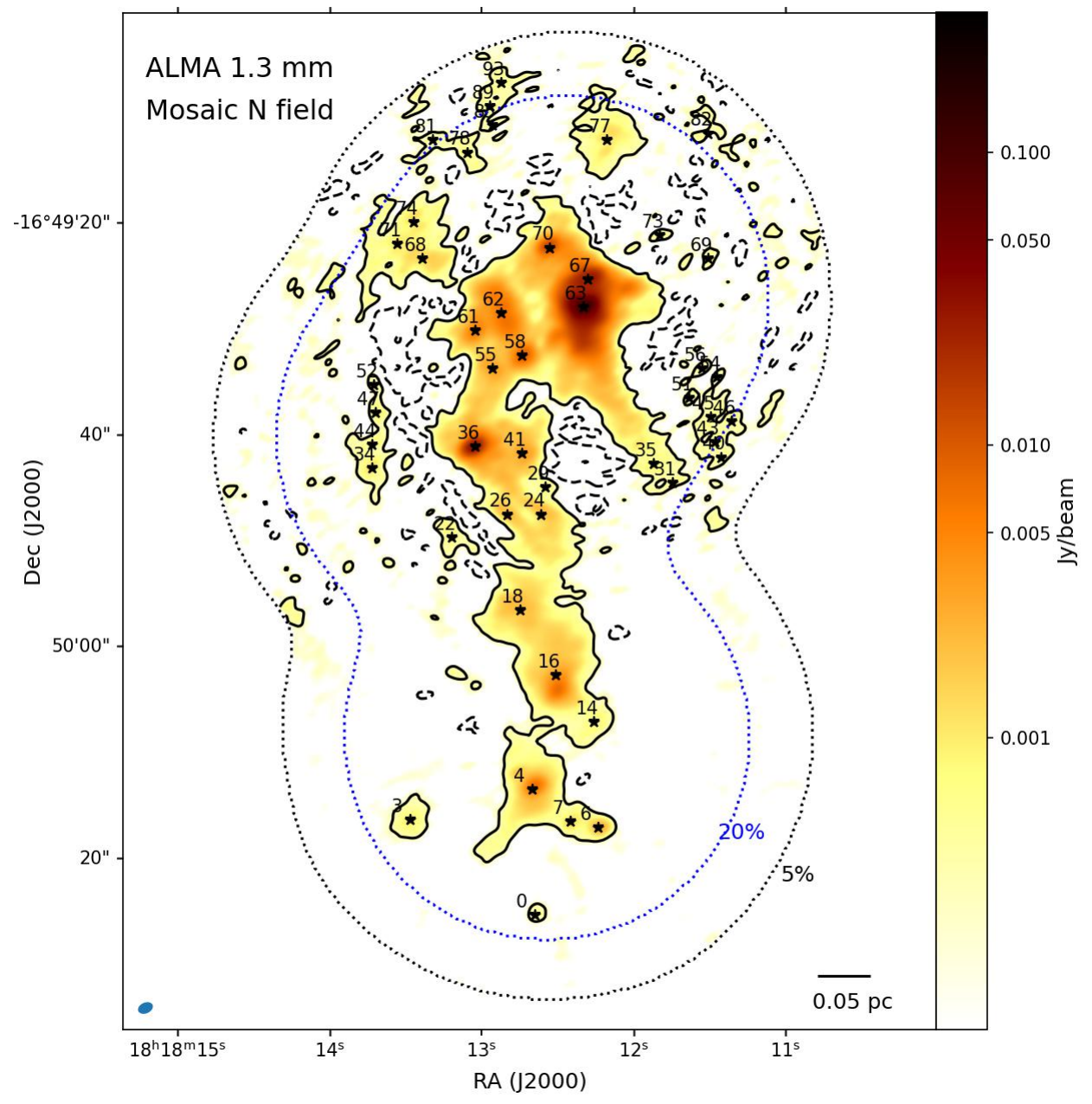}
\includegraphics[width=9.5cm]{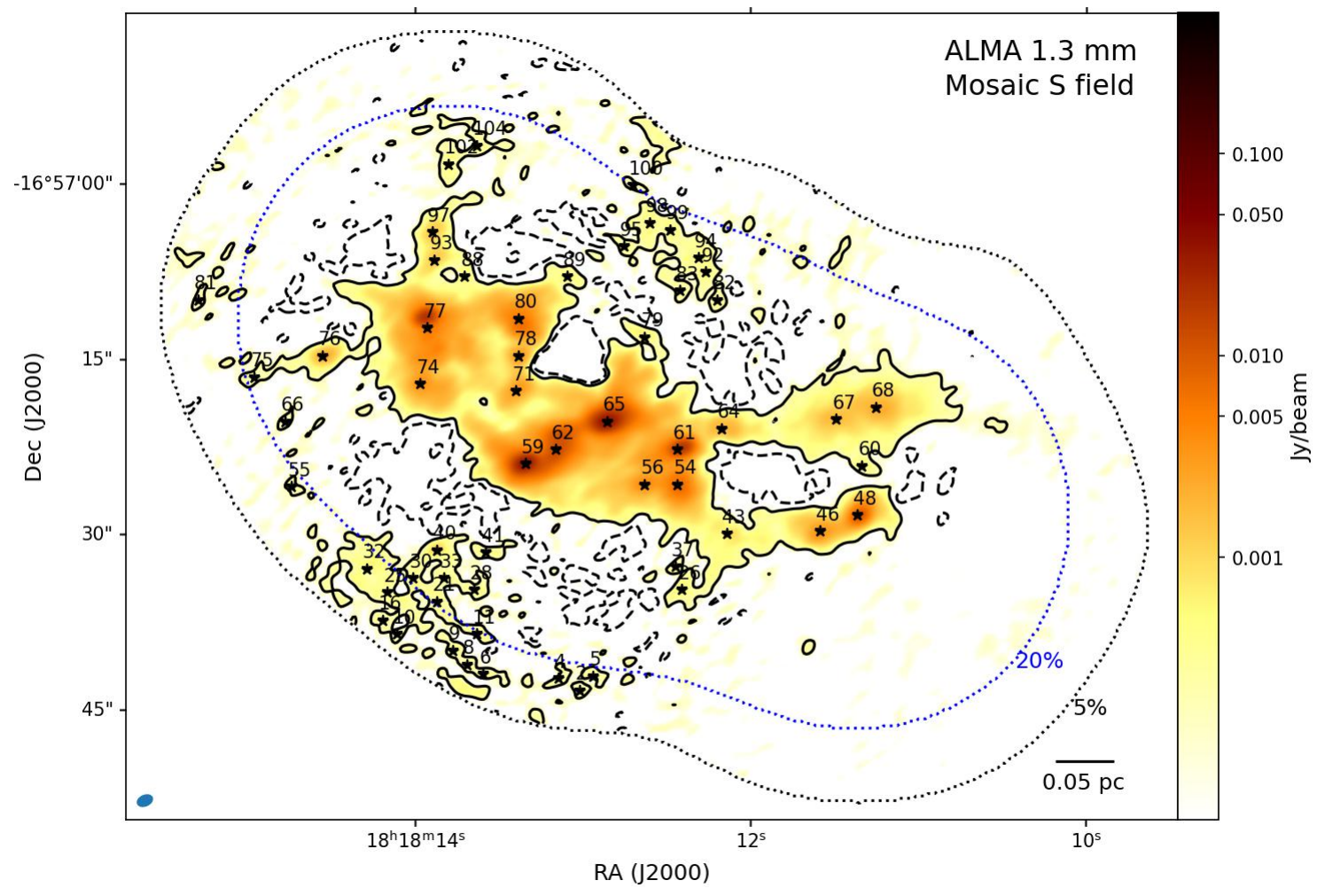}
\caption{Continuum emission in the field of N.4, N.5, N.7, mosaic N, and mosaic S. Maps are plotted in log-scale.\\
Contour levels for N.4, N.5 and N.7, mosaic N, and mosaic S are -3$\sigma$ (dashed) and 3$\sigma$ (solid), with the rms noise level $\sigma$=0.10, 0.05, 0.05, 0.10 and 0.06 mJy~beam$^{-1}$. 
Star signs denote the center position of {\fontfamily{qcr}\selectfont leaf} structure identified in the dendrogram. Synthesized beam sizes as an ellipse are marked at the bottom-left corner of the panel. Dotted contours illustrate the coverage of each field, plotted at the sensitivity level of 5\% (black) and 20\% (blue).}
\label{fig:continuum_emission}
\end{figure*}

The continuum map with the threshold contour levels and the location of the {\fontfamily{qcr}\selectfont leaf} structures of the northern and southern fields identified from the dendrograms are shown in Figure \ref{fig:continuum_emission}. Due to the effect of the primary beam correction, structures at the edge of the field with a peak emission smaller than 3$\sigma_{local}$ (their nearby noise level) were not included as detection. In addition to structures at the edge of the primary beam, 14 structures in the N mosaic field and 6 structures in the S field are rejected from the dendrogram statistics. We identified 221 structures in total, including 126 {\fontfamily{qcr}\selectfont leaf} structures. Compared with the position of the sources identified from the 3 mm observations in \cite{Ohashi_2016}, the {\fontfamily{qcr}\selectfont leaf} structures identified in the 1.3 mm emission are in a general agreement under a 1.5$''$ separation criterion with those in the 3 mm. With a higher sensitivity and higher angular resolution in the 1.3 mm observations, some condensations were detected from which single sources were seen in the 3 mm continuum, while there are also a few sources undetected in the 1.3 mm band due to its limited field of view.

\subsection{Dense Core Properties}
Dendrogram statistics, including the flux, radius, position, and size of each structure, are derived from the {\fontfamily{qcr}\selectfont Astrodendro} package. The 1.3 mm emission arises mainly from the dust emission \citep{Ohashi_2016}, and the masses of the core structures can be computed with 

\begin{equation}
    M_{gas} = \frac{F_{\nu}d^2}{\kappa_{\nu} B_{\nu}(T_d)}\eta,
\end{equation}
where $F_{\nu}$ is the flux observed at the frequency $\nu$, $d$ denotes the distance to the target source, $\eta$ is the dust-to-gas mass ratio (assumed to be 100, \cite{Hildebrand_1983}), $\kappa_{\nu}$ is the dust opacity at $\nu$, and $B_{\nu}$ is the Planck function at a dust temperature $T_d$ and frequency $\nu$. Assume dust grains with thin ice mantles and gas density of 10$^6$ cm$^{-3}$ \citep{Ossenkopf_1994}, dust opacity $\kappa_{1.3mm}$ = 0.9 cm$^2$ g$^{-1}$ is adopted. The gas temperatures of the core structures are approximated by the kinematic temperature of ammonia from the NH$_3$ (J, K) = (1, 1) line emission computed by \citet{Busquet_2013} with the Very Large Array and the Effelsberg 100 m telescope. The average values of their neighboring field are adopted as an approximation for structures with no corresponding temperature at their positions. For example, the averaged temperature value 14.56 K of the {\fontfamily{qcr}\selectfont trunk} structure in the N.5 field is used for all the structures in the N.7 field where temperature estimates from NH$_3$ are not available. We derived masses of {\fontfamily{qcr}\selectfont leaf} structures ranging from $\sim$0.1 $M_{\odot}$ to 20.4 $M_{\odot}$, while for the {\fontfamily{qcr}\selectfont trunk} structure, the field 4, 5 and 7 harbor the masses of $\sim$7.6-12.8 $M_{\odot}$, much smaller than the mosaic field N and S, in which an estimated total mass of 112.0 $M_{\odot}$ and 74.0 $M_{\odot}$ are identified from the dendrogram. The 1$\sigma$ mass sensitivity is $\sim$0.007 $M_{\odot}$ when adopting an rms level of 1$\times$10$^{-4}$ Jy~beam$^{-1}$ and an average temperature of 18 K. According to \citet{Li_2020}, the uncertainty of the gas mass calculation could reach 57$\%$ after taking into account of uncertainties in the distance of the cloud, NH$_3$ temperatures, the dust emissivity, the dust-to-gas ratio, etc. Column densities are evaluated by assuming a projected area of $\pi r^2$ and thus density = $M_{gas}/(\mu m_{H} \pi r^2)$. Densities vary from 6.3$\times$10$^{21}$ cm$^{-2}$ to 2.2$\times$10$^{24}$ cm$^{-2}$, with a mean and median value being 8.5$\times$10$^{22}$ cm$^{-2}$ and 4.4$\times$10$^{22}$ cm$^{-2}$, respectively. Physical parameters of the core structure for the N.4, N.5, N.7, N mosaic, and S mosaic field are summarized in Table \ref{tab:continuum_phy_para}.

We derived virial parameters and Mach numbers to study dynamical properties of these core structures. By assuming uniform density structures, the virial mass is calculated by 

\begin{equation}
    M_{vir} = 210 \times (\frac{r}{1\ pc}) \times (\frac{\Delta v}{1 \ km \ s^{-1}})^2 \ M_{\odot},
\end{equation}
where $r$ is the radius in FWHM and $\Delta v$ is the FWHM line width ($\Delta v=2\sqrt{2ln(2)}\sigma_V$, where $\sigma_V$ is the velocity dispersion) \citep{Ohashi_2016}. Then the virial parameter $\alpha_{vir}$ is the ratio.

\begin{equation}
    \alpha_{vir} = \frac{M_{vir}}{M_{gas}}.
\end{equation}

The line width $\Delta v$ is derived from the velocity dispersion $\sigma_V$ of the averaged N$_2$D$^+$ J = 3-2 emission spectrum over the core structures using the data from the same ALMA observations with the CO 2-1 and 1.3 mm dust emission. We performed 1-dimensional Gaussian fits to the spectrum using astropy.modeling package and derived the dispersion $\sigma_{obs}$. For structures with multiple velocity components, their spectra are fitted by the sum of Gaussian functions for each of the components. To account for the effect of broadening due to the spectral resolution, we use the following approximation 

\begin{equation}
    \sigma_V=\sqrt{\sigma_{obs}^2 -\Delta_{ch}^2/(2\sqrt{2ln2})^2},    
\end{equation}
where $\sigma_{obs}$ is the observed velocity dispersion towards the core structures measured from the N$_2$D$^+$ spectra, and $\Delta_{ch}=$ 0.3 km~s$^{-1}$ is the velocity channel width. For structures with a fitted velocity dispersion that is smaller than the dispersion caused by channel width ($\sim$ 0.12 km~s$^{-1}$), $\sigma_{obs}$ is directly used as an overestimation. The average and median line width $\Delta v$ of all the identified structures are 0.96 km~s$^{-1}$ and 0.87 km~s$^{-1}$.

With the linewidths from Gaussian fitting, the nonthermal velocity dispersion is obtained by 

\begin{equation}
    \sigma_{nt}=\sqrt{\sigma_V^2-\sigma_{th}^2},
\end{equation}
where $\sigma_{th}$ is the thermal velocity dispersion with $\sigma_{th}=\sqrt{(k_B T)/(\mu m_H)}=9.08\times 10^{-2}$ km s$^{-1}(\frac{T}{K})^{0.5}\mu^{-0.5}$, where $k_B$ is the Boltzmann constant, $T$ is the gas temperature estimated from NH$_3$, $\mu=m/m_H$ is the molecular weight, and $m_H$ is the mass of hydrogen atom. Mach number is calculated by $\mathcal{M}=\sqrt{3}\sigma_{nt}/c_s$, where the sound speed $c_s$ is the thermal dispersion of gas of mean particle mass estimated as 2.37$m_H$ \citep{Kauffmann_2008}. Overall statistics show a strong tendency of transonic and supersonic motions with a mean Mach number of 2.8 and median of 2.7, while fields N.5 and N.7 are less turbulent (mean Mach number: 1.9, median Mach number: 2.0) compared to the other three fields (mean Mach number: 2.9, median Mach number: 2.7). Dynamical parameters of the core structure for the N.4, N.5, N.7, N mosaic, and the S mosaic field are summarized in Table \ref{tab:continuum_dyn_para}.

\subsection{Molecular Outflows}
\begin{figure*}[h!]
\centering
\includegraphics[width=13cm]{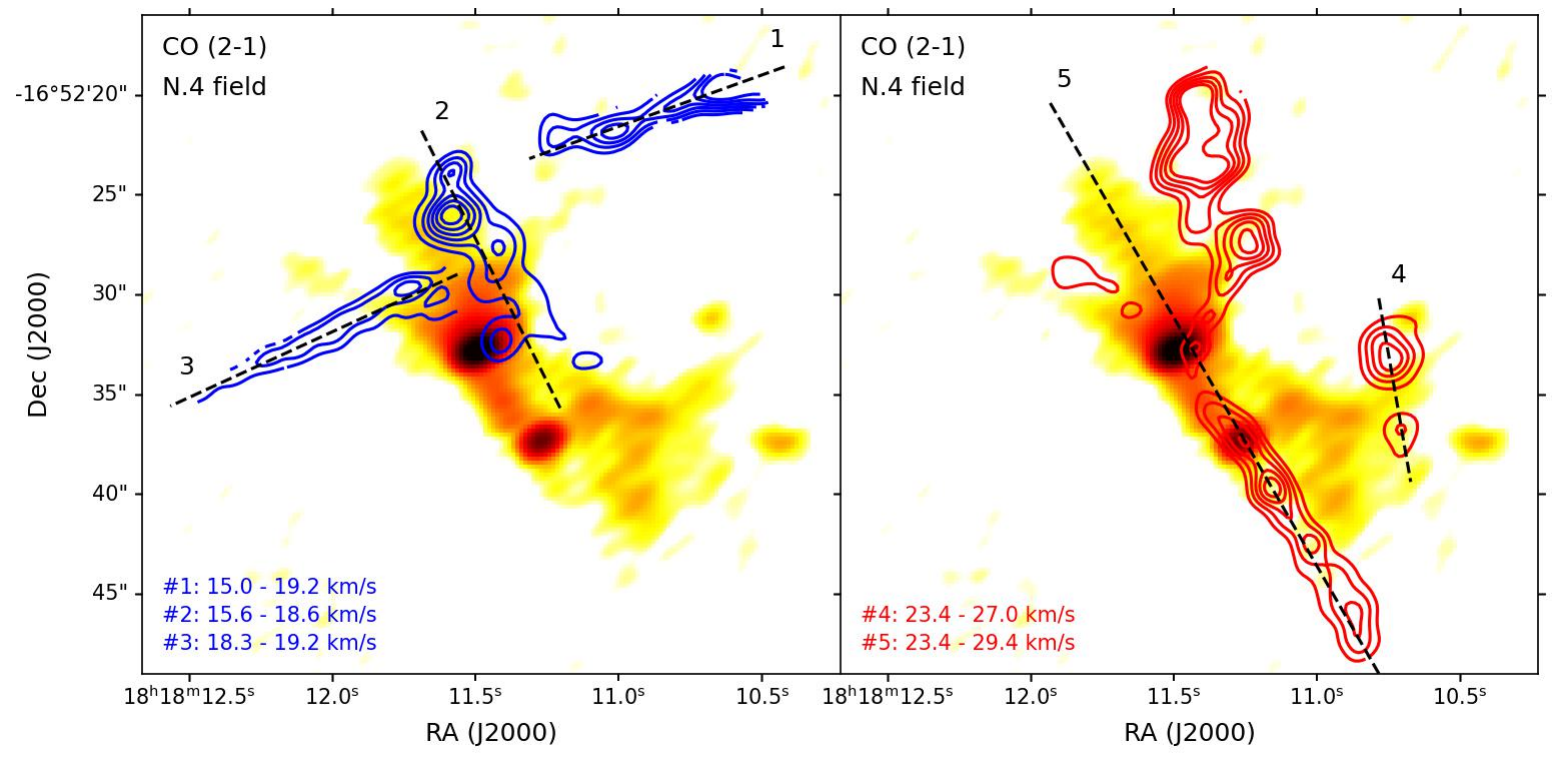}
\includegraphics[width=13cm]{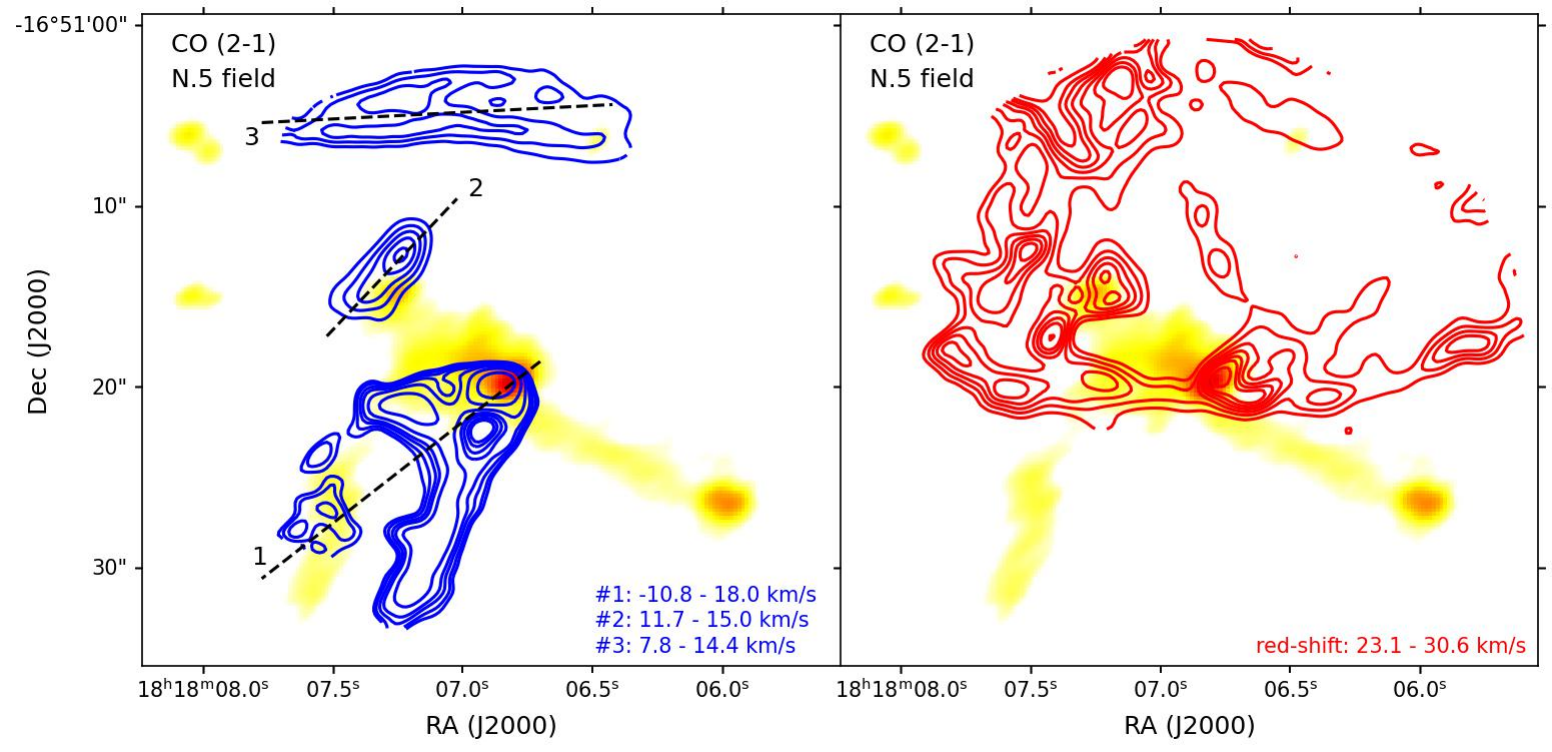}
\includegraphics[width=8cm]{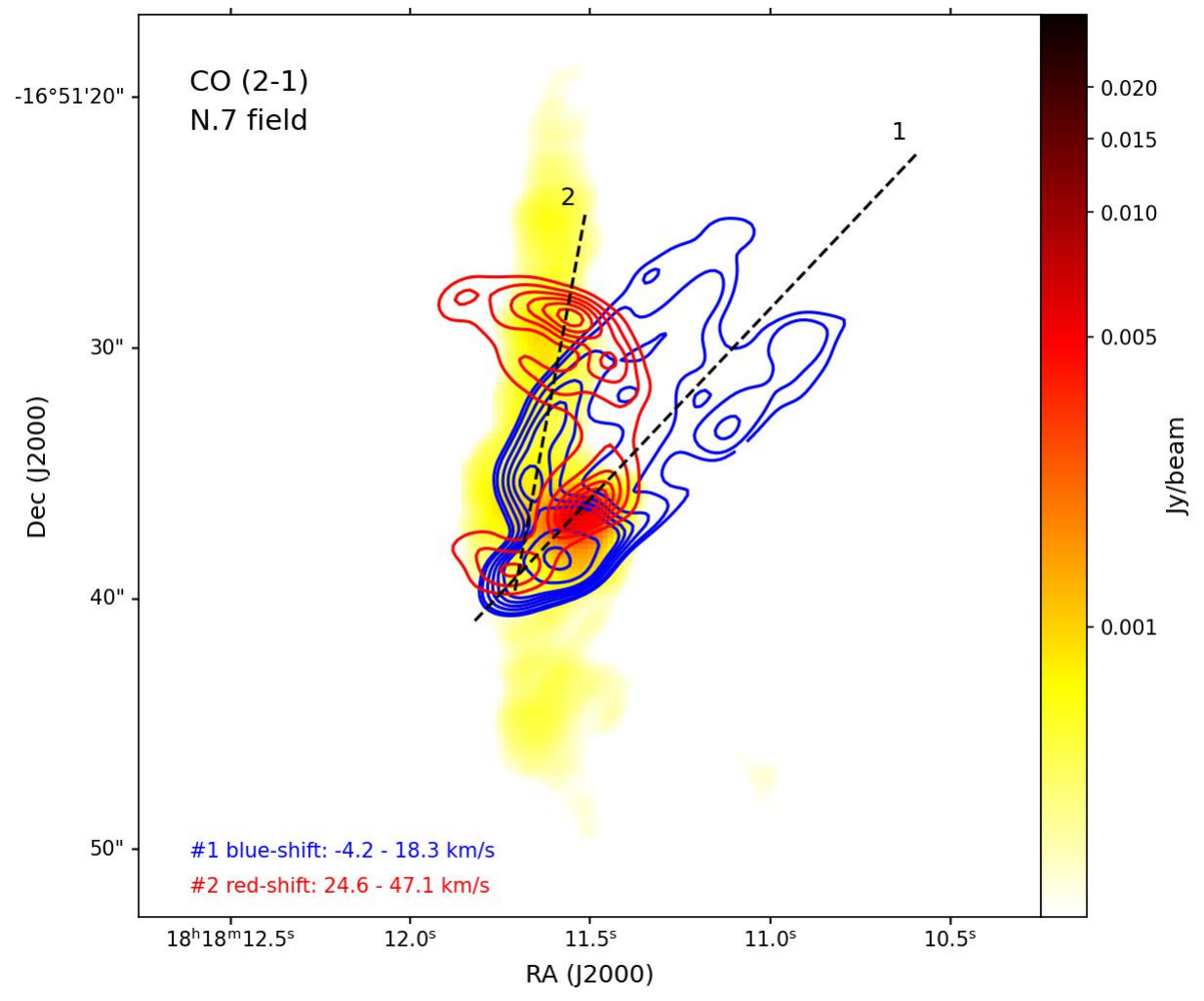}

\caption{Integrated intensities of the CO 2-1 emission in blue and red contours tracing the outflows in the single field of N.4 (upper panel), N.5 (middle panel), and N.7 (lower panel). The 1.3 mm continuum emission maps are overlaid with color map in logarithmic scales with emission levels from 2$\times$10$^{-4}$ to 3$\times$10$^{-2}$ Jy~beam$^{-1}$. Integrated velocity range of each outflow is at the bottom corner of each subplot.}
\label{fig:n_sig_outflow}
\end{figure*}

\begin{figure*}[h!]
\centering
\includegraphics[width=10.4cm]{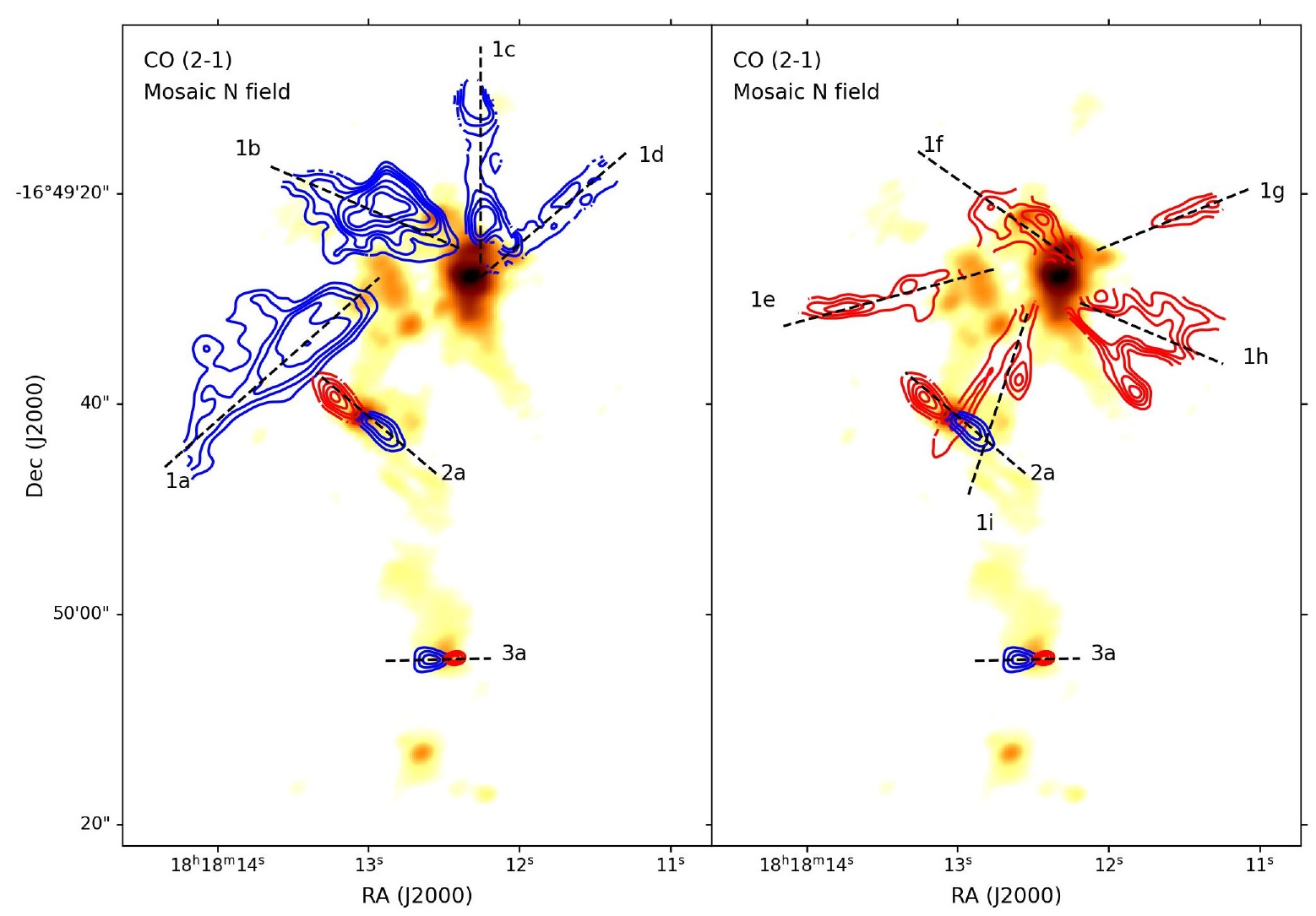}
\includegraphics[width=5.88cm]{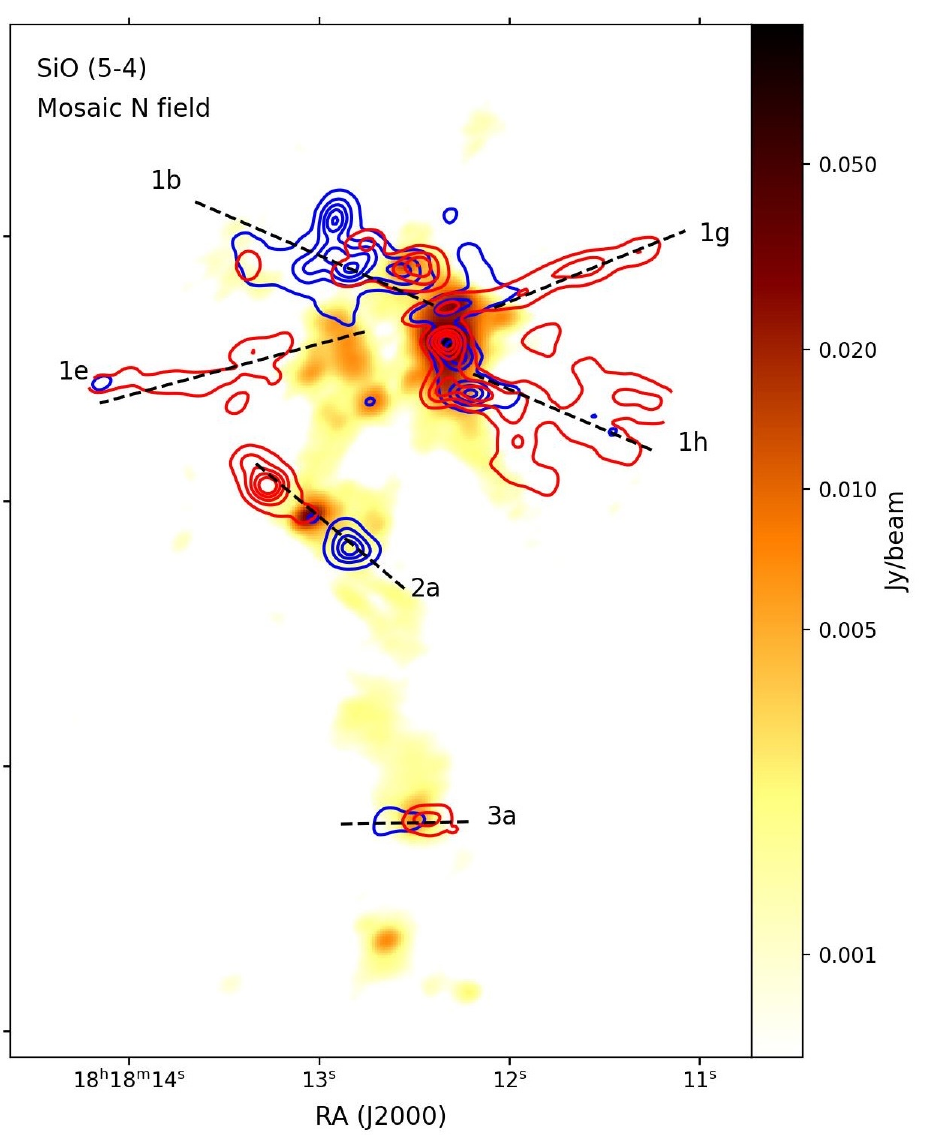}
\includegraphics[width=8.0cm]{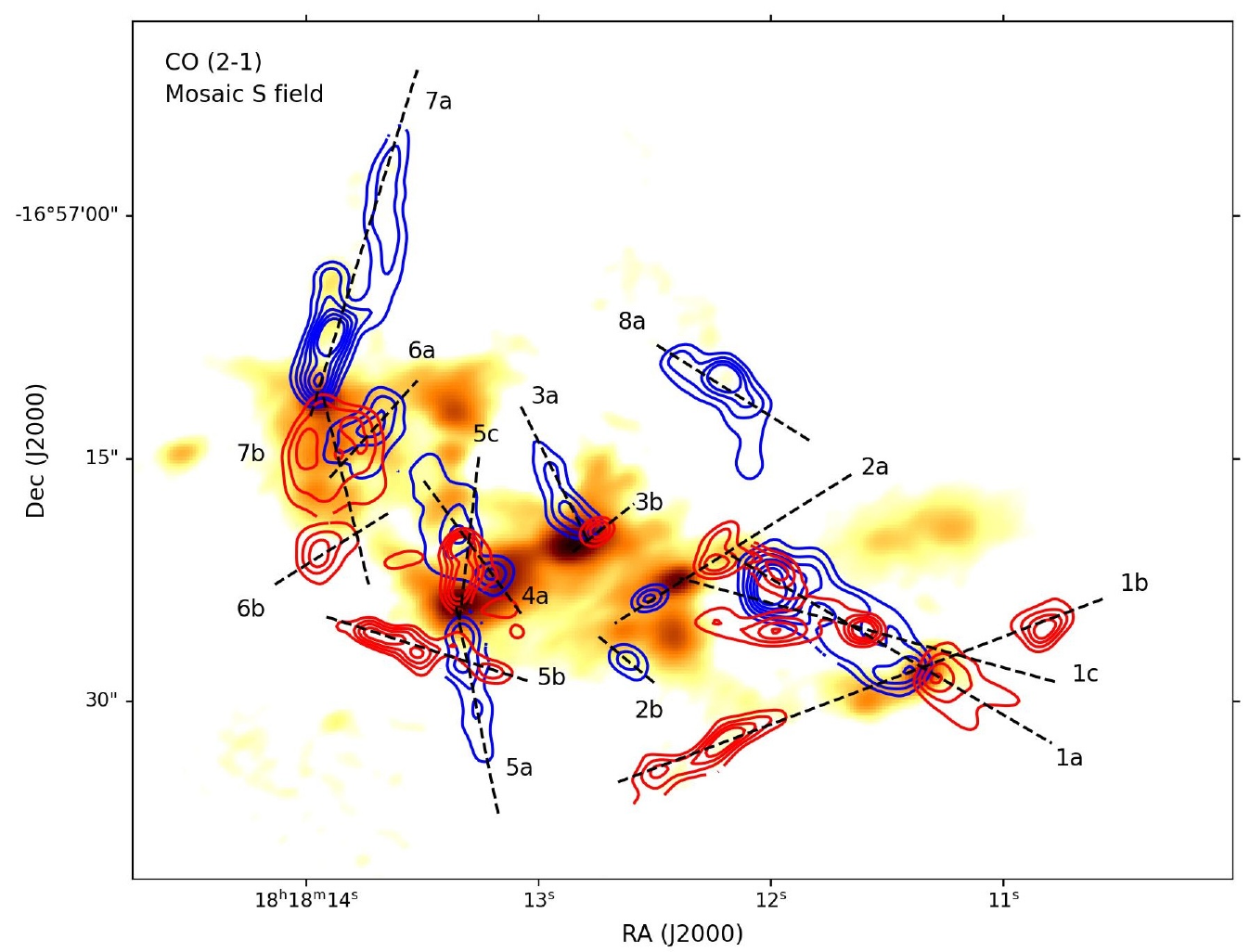}
\includegraphics[width=8.02cm]{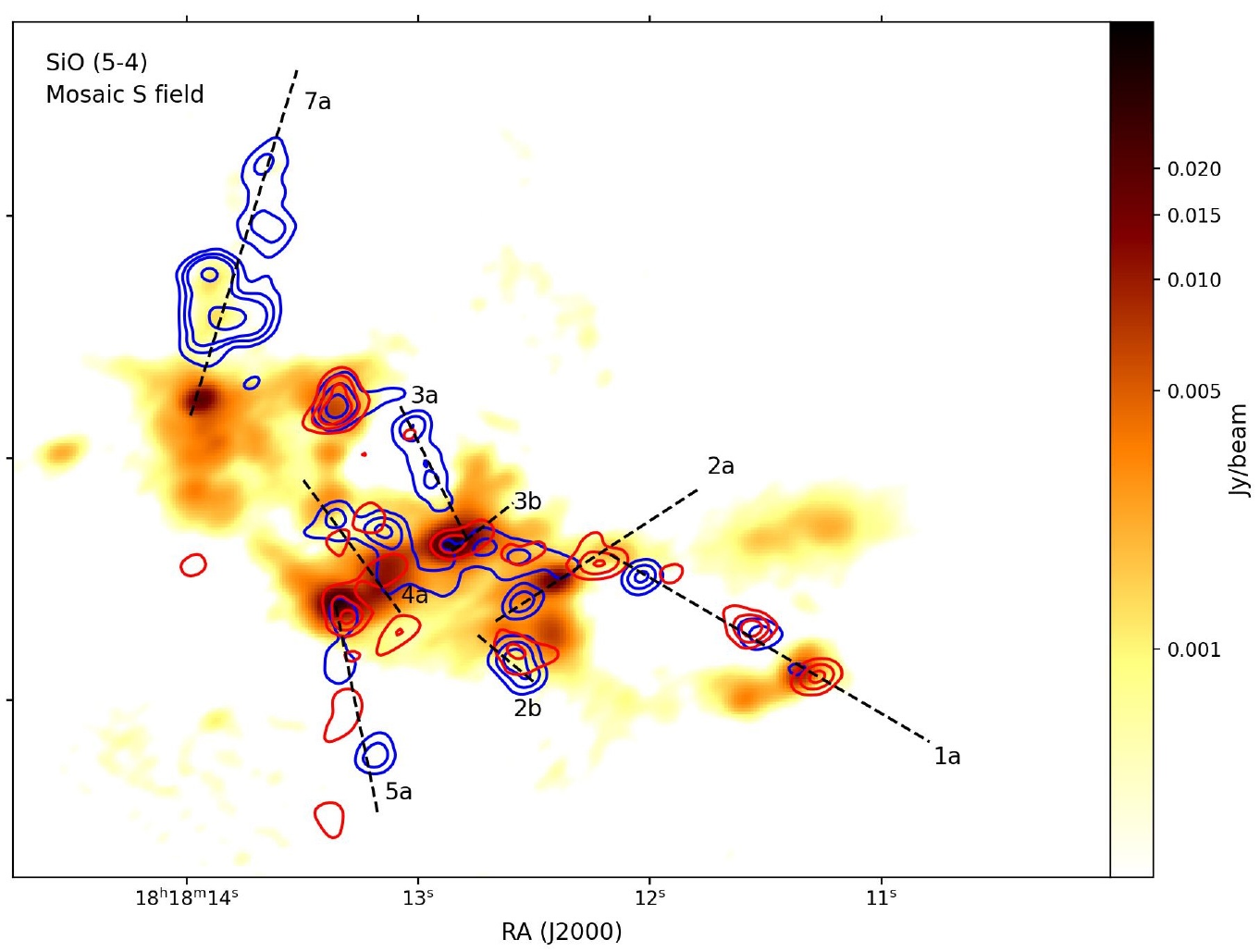}
\caption{Integrated intensity map of the CO 2-1 and SiO 5-4 emission tracing the outflows in the Northern (upper panel) and Southern (lower panel) mosaic field. 1.3 mm continuum emission is overlaid with color map in log-scale. Integrated velocity range of each outflow is tabulated in Table \ref{tab:outflow_parameter}.}
\label{fig:mosaic_outflow}
\end{figure*}

Outflows are detected in multiple tracers such as in the CO, SiO, and CH$_3$OH emission. Outflow morphologies indicated by the CO 2-1 and SiO 5-4 line emission are presented in Figure \ref{fig:n_sig_outflow} and Figure \ref{fig:mosaic_outflow}. In each field, there are outflows extending beyond the peak continuum emission, which is identified as dense cores. We find spatially extended outflows, such as outflow 5 in the N.4 field, and high-velocity outflows with emission extended across a wide range of velocity channels, such as outflow 1 in the southeast of the N.5 field. From the blue-shifted panel in fields N.4 and N.5, outflows such as $\#$1 and $\#$3 in N.4 and $\#$3 in N.5 may originate outside of the pointing coverage. The red-shift outflow in the N.5 field exhibits a diffuse CO emission extended beyond the edge of the field. In the N.7 field, strong blue-shifted outflow emission exhibits rich structures shown from the contours. In the red-shift channels, the bend in the CO emission may require more sensitive observations to identify its powering source. In the northern mosaic field, in addition to the bipolar outflows 2a and 3a found in both the CO 2-1 and SiO 5-4 emission, complicated and radial-shaped outflows are widespread in the upper region. In the southern fields, however, more bipolar and elongated features are shown. 

Physical parameters, including mass, momentum, and energy, are derived following the formulation in \cite{Zhang_2005}. We compute the column density with
\begin{equation}
    N_{CO}=1.08\times exp(\frac{16.59}{T_{ex}}) \cdot (T_{ex}+0.92) \int T_B dv \cdot \frac{\tau}{1-e^{-\tau}}
\end{equation}
following \cite{Garden_1991}. Then the outflow mass is calculated by assuming the CO emission to be optically thin and the CO abundance to be [CO/H$_2$] = 10$^{-4}$. The excitation temperature is approximated by the brightness temperature of the CO emission at the peak emission of the outflow. The projection effect of outflows is ignored in the computation. The outflow mass ranges from $\sim$ 2$\times$10$^{-3}$ $M_{\odot}$ to 0.3 $M_{\odot}$, and the typical momentum and energy are in the order of $\sim$ 10$^{-2}-$10$^{-1}$ $M_{\odot}$ km s$^{-1}$ and $\sim$ 10$^{-1}-$10$^{0}$ $M_{\odot}$ km$^2$ s$^{-2}$. Statistics of the outflow parameters are tabulated in Table \ref{tab:outflow_parameter}. 

\section{Discussion} \label{sec:discussion}
\subsection{Dynamical States and Motions of Dense Core Structures}

\subsubsection{Virial Properties of Dense Cores}
The median value of virial parameters in dense cores identified in this study is 1.4. Without the external pressure, around 68$\%$ of the structures are gravitationally bound ($\alpha_{vir} < 2$) and 36.3$\%$ of the total structures are gravitationally unstable ($\alpha_{vir} < 1$). There is an anti-correlation between the virial parameter and the column density shown in Figure \ref{fig:viri_para_den}. $\alpha_{vir}$ decreases with increasing densities, implying that denser regions tend to be more gravitationally unstable. Protostellar cores associated with outflows tend to have higher column densities compared to the median density of all dendrogram structures, with $>6\times 10^{22}$ cm$^{-2}$ and mainly at $10^{23}$ cm$^{-2}$. These dense cores have low virial parameters and are mostly gravitationally unstable. These characteristics are consistent with surveys of high-mass clumps by \cite{Li_2023} who reported that virial parameters decrease from presteller to protostellar cores. Combining data of parsec-scale filaments and clumps, and massive dense cores, \cite{Chen_2019} found an anti-correlation between $\alpha_{vir}$ and spatial scales in G14, which implies that the effect of gravity becomes stronger compared to turbulence in dense cores. For structures that are gravitationally stable, it is possible that the overall non-thermal motion provides sufficient support against the pull by gravity, but their substructures at smaller spatial scales experience gravitational collapse. In this analysis, the effects of the magnetic field and the external pressure are not included in the calculation. While the magnetic field provides internal support to the core structure, the external pressure exerts a force that helps confine the core.

\subsubsection{Supersonic Motion in Dense cores}
\begin{figure*}[h!]
\includegraphics[width=5.9cm]{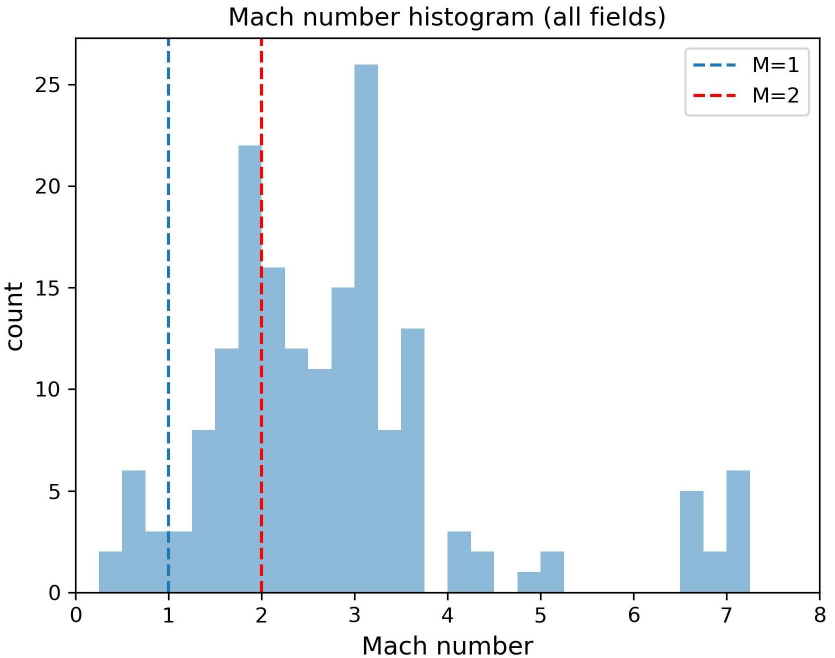}
\includegraphics[width=5.9cm]{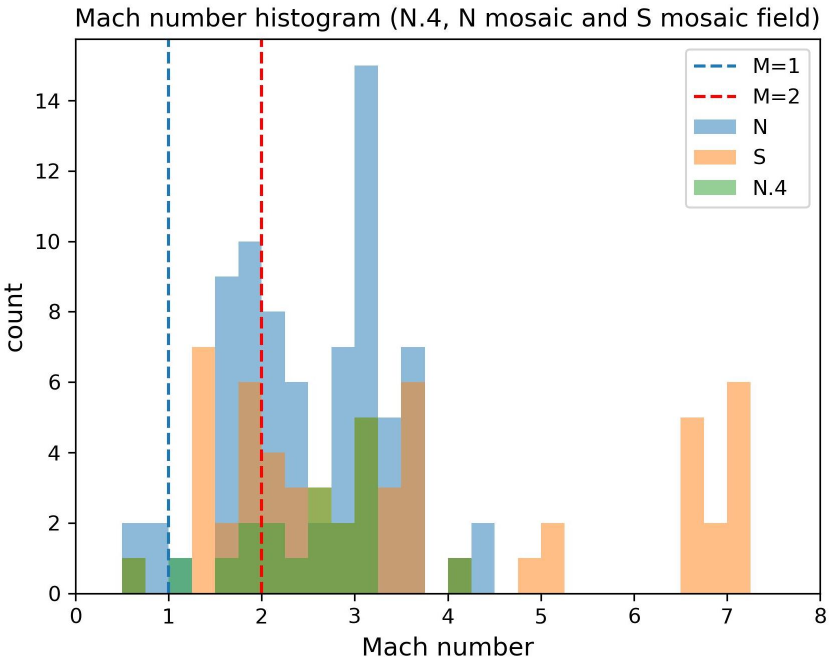}
\includegraphics[width=5.9cm]{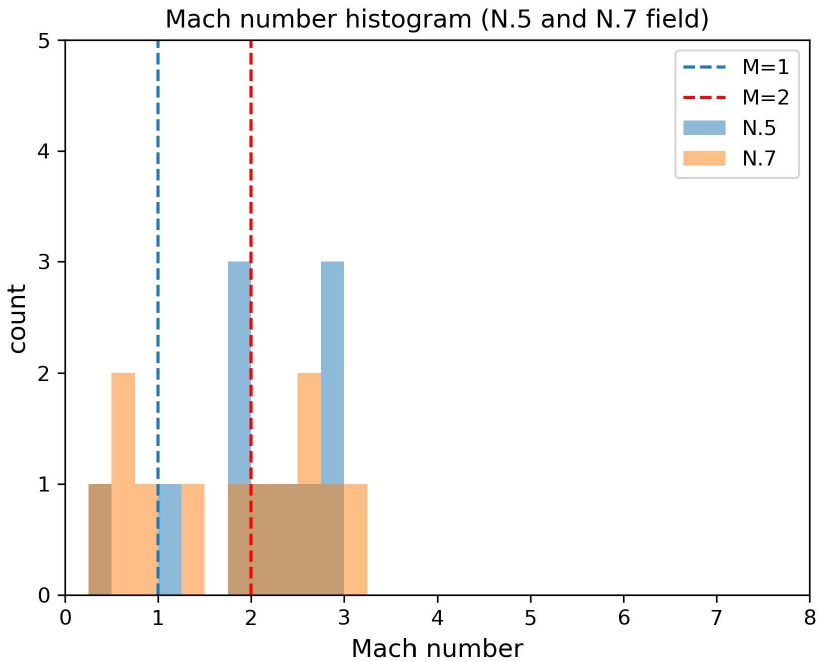}
\caption{Histograms of Mach numbers in different observing regions. The left plot includes the structures from all of the mosaic and single fields. The middle plot shows the Mach number of structures in the mosaic N, mosaic S, and N.4 fields, and the right panel presents the Mach number in the field of N.5 and N.7. Vertical lines $\mathcal{M}$=1 and $\mathcal{M}$=2 denote the boundary between subsonic to transonic and transonic to supersonic motion.}
\label{fig:mach_hist}
\end{figure*}

We found that most dense cores in the observed fields are supersonic, with 68.5$\%$ of them having Mach number $\mathcal{M}>2$ as shown in Figure \ref{fig:mach_hist}. Among the fields observed, the N.4 and mosaic N and S fields exhibit mostly supersonic motions ($\sim$70$\%$), while in the N.5 and N.7 fields, supersonic turbulence accounts for smaller proportions ($\gtrsim$50$\%$) of the core structures, as shown by the statistics summarized in Table \ref{tab:mach_num_stat}. Overall, about 93.8$\%$ of the structures exhibit $\mathcal{M} \geq$ 1, indicating a transonic-to-supersonic nature of the turbulence. This is in contrast with studies by \citet{Li_2023, Li_2020} who reported subsonic or transonic turbulence in an infrared dark cloud in NGC~6334. The supersonic motions in regions of the N.4, mosaic N and S fields may arise from star formation activities compared to the fields N.5 and N.7 where more structures exhibit subsonic and transonic motions. Moderate transonic non-thermal motions of the G14.225 filaments are reported by \citet{Chen_2019}. At the core scale, gravitational collapse in dense cores and protostellar outflows may broaden the line width \citep{Li_2023}. In addition, other non-turbulent and non-thermal motions such as rotation or infall can also affect line widths and support a structure against gravitational collapse. 

\begin{figure*}[h!]
\centering
\includegraphics[width=8.45cm]{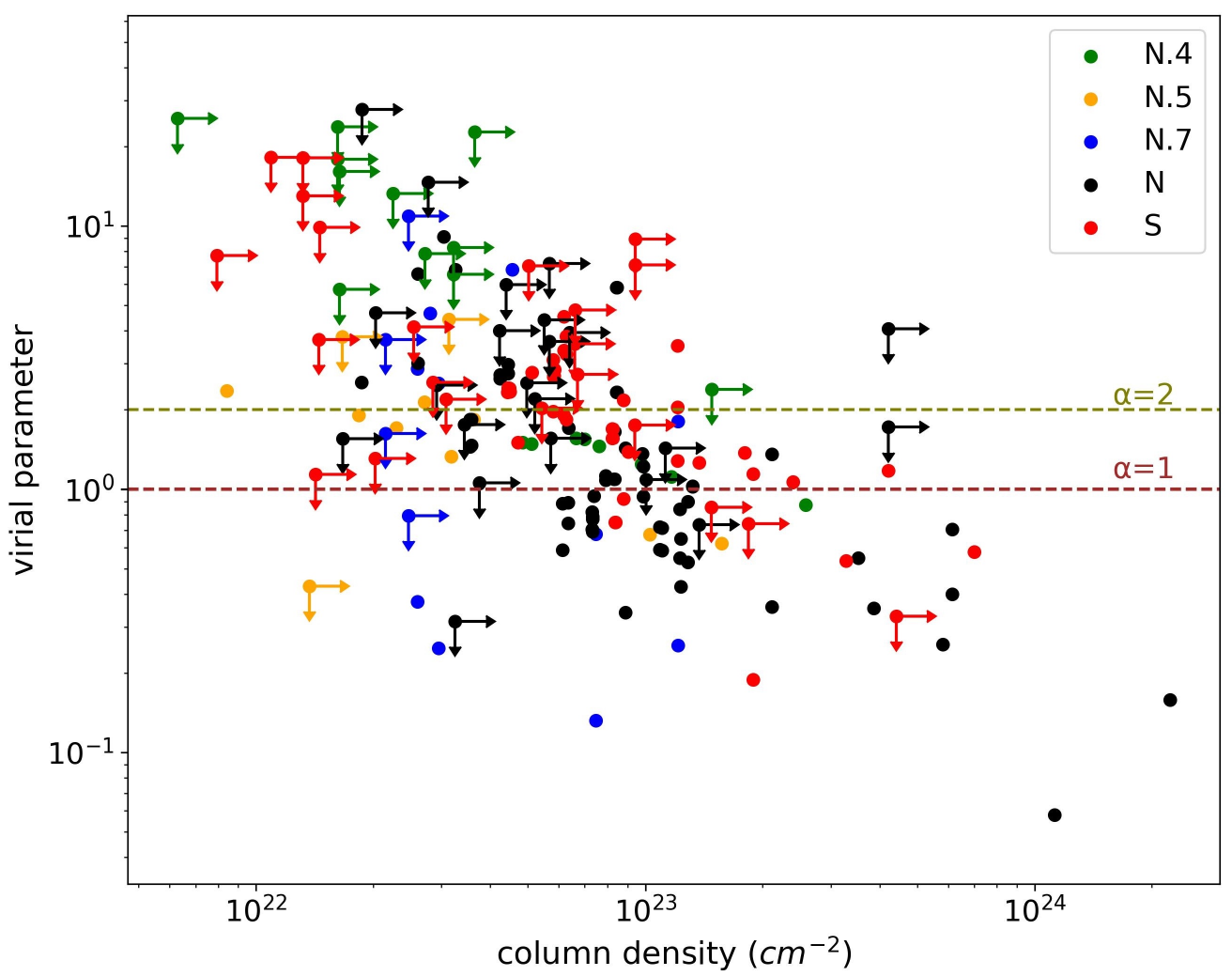}
\includegraphics[width=8.55cm]{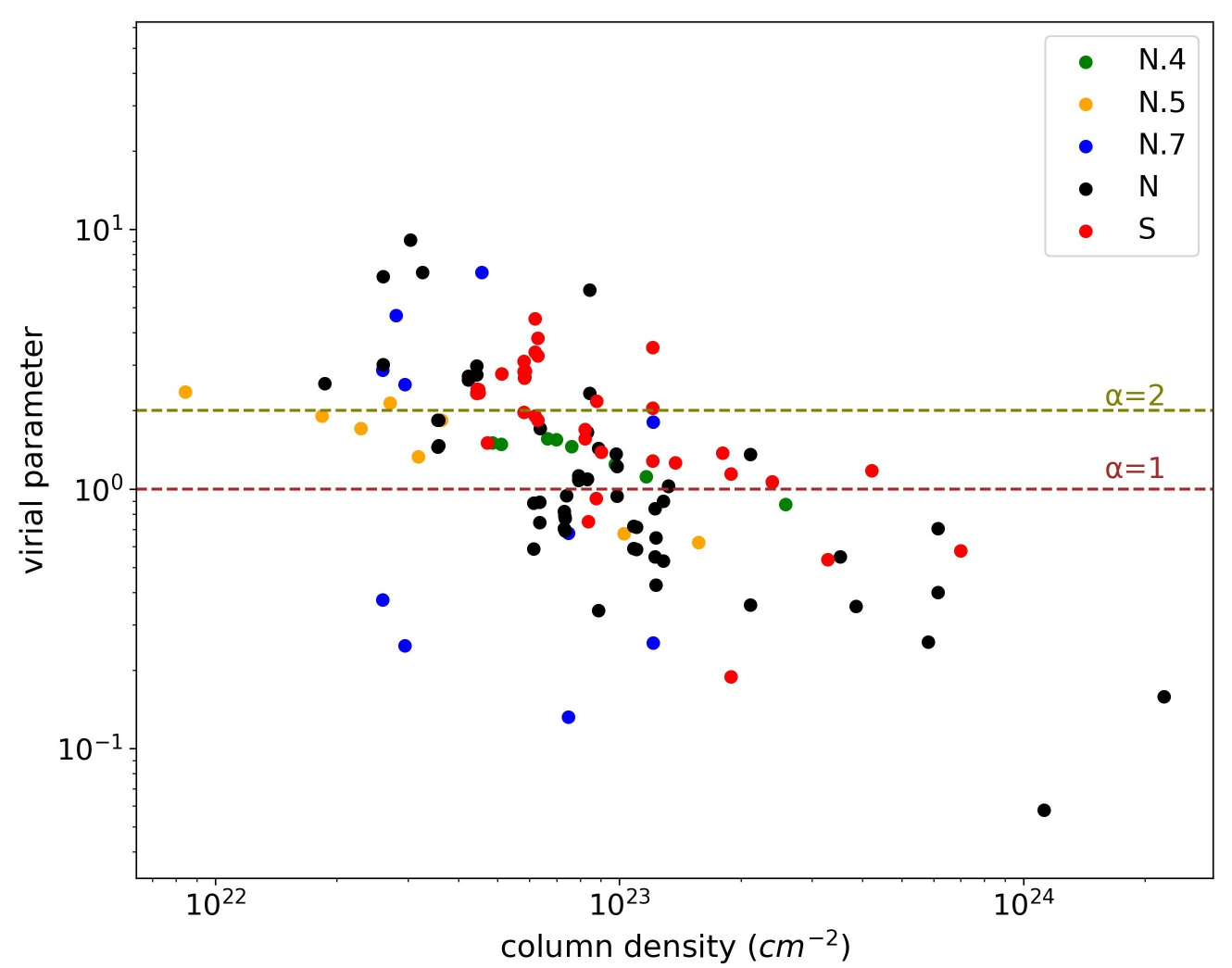}
\caption{Scatter plot of virial parameters versus column densities. Left panel shows all data points including those with unresolved spatial sizes, and right panel shows the resolved dendrogram structures deconvolved with the beam size. Arrows in the left panel indicate that for overestimated spatial sizes in the {\fontfamily{qcr}\selectfont leaf} structures, the column density serves as a lower limit while the virial parameter is an upper-limit estimation.}
\label{fig:viri_para_den}
\end{figure*}

\begin{table*}[h!]
\caption{Mach Number Statistics in observed fields}
\label{tab:mach_num_stat}
\centering
\begin{tabular}{|c|c|c|c|c|c|c|}
\hline
Field & $\mathcal{M}$ & $\mathcal{M}_{mean}$ & $\mathcal{M}_{median}$ & $\mathcal{M}\leq 1$ & $1<\mathcal{M}\leq2$ & $\mathcal{M}\geq 2$ \\ 
\hline
All fields & 0.28 $\sim$ 7.21 & 2.80 & 2.67 & 6.2\%         & 25.3\%  & 68.5\%    \\ \hline
N.4        & 0.61 $\sim$ 4.20 & 2.48 & 2.65 & 5.3\%         & 21.0\%  & 73.7\%   \\ \hline
N.5        & 0.28 $\sim$ 2.82 & 2.06 & 2.00 & 9.1\%         & 36.4\%  & 54.5\%   \\ \hline
N.7        & 0.40 $\sim$ 3.03 & 1.78 & 1.93 & 33.3\%        & 16.7\%  & 50.0\%   \\ \hline
N mosaic   & 0.67 $\sim$ 4.31 & 2.53 & 2.57 & 5.2\%         & 26.0\%  & 68.8\%   \\ \hline
S mosaic   & 0.67 $\sim$ 7.21 & 3.58 & 3.14 & 1.7\%         & 25.4\%  & 72.9\%   \\ \hline
\end{tabular}
\end{table*}

The analysis presented here can be affected by the limited spatial and spectral resolutions. About 48$\%$ of the {\fontfamily{qcr}\selectfont leaf} structures with small sizes identified by the Dendrogram are not deconvolved with the synthesized beam. Thus, for those structures, the virial masses calculated from the sizes would be systematically overestimated and shown as upper limits in Figure \ref{fig:viri_para_den}, while their column densities are underestimated. Line width measurements may also be affected and thus broadened by multiple factors. There could be a blending of multiple velocity components within the identified structures that are not clearly separated in frequencies. Due to a limited spectral resolution of the N$_2$D$^+$ data (velocity channel width of $\sim$ 0.3 km~s$^{-1}$), line widths smaller than the channel width are overestimated. With limited signal-to-noise ratios in the satellite components, the spectra of the N$_2$D$^+$ 3-2 line emission are fitted by a Gaussian rather than using the hyperfine-component profile, which leads to larger velocity dispersions than fiting hyperfine components. The velocity dispersion differences between Gaussian fitting and hyperfine-components fitting can be $\sim$ 0.07$-$0.10 km~s$^{-1}$. Therefore, the actual motions in some cores could be smaller, and the cores could be gravitationally unstable.

\subsubsection{Fragmentation in Dense Cores}
The analysis of the continuum emission identified dense cores with a range of masses. The typical thermal Jeans mass in the clumps is $\sim$ 2 M$_\odot$ by assuming a density of 5$\times$10$^4$ cm$^{-3}$ and an average temperature of 15 K. However, there exist dense cores with  masses about 10 times greater than the Jeans mass, yet the core mass is approximately comparable to the Jeans mass if the sound speed is replaced by the observed velocity dispersion. From the hierarchical dendrogram analysis, the spatial scale of {\fontfamily{qcr}\selectfont leaf} structures tabulated in Table \ref{tab:continuum_phy_para} is $\lesssim$ 0.01 pc, implying that the parent {\fontfamily{qcr}\selectfont trunk} structures at $\sim$ 0.05 $-$ 0.15 pc have fragmented into substructures. Some of these substructures, or condensations, are starless or still in dynamical equilibrium and prestellar stage; Others harbor protostars, as demonstrated by associations with molecular outflows shown in Figures \ref{fig:n_sig_outflow}, \ref{fig:mosaic_outflow} and emission in X-ray or radio wavelengths shown in Figure \ref{fig:multi-wavelength}.

Theoretical models on core accretion of massive stars offer possible interpretations and exhibit limitations when compared with the observations. The turbulence accretion model proposed by \cite{Mckee_2002} suggested that massive cores monolithically collapse to form stars. The foundational assumption is the self-similar, self-gravitating, and approximately hydrostatic equilibrium features of dense cores. Under a high-pressure environment, supersonic turbulence dominates massive cores that have masses that far exceed the thermal Jeans mass. Another scenario is the competitive accretion model by \cite{Bonnell_2002} which considers massive star formation as part of a protostellar cluster formation. The scenario starts with a distributed mass fragments at approximately thermal Jean mass. A non-uniform accretion rate across the cluster gives rise to protostars with a range of stellar masses, among which massive stars are found at the cluster center. Observational results in G14.225 exhibit supersonic turbulence in the massive cores and show the existence of cores that are significantly above the thermal Jeans mass, which seems to favor the turbulent accretion model. However, fragmentation in core and condensation scales is inconsistent with the monolithic collapse proposal. As outflow detection might reveal ongoing accretion, and the maximum core mass identified from the study is 20.4 $M_{\odot}$. The lack of high-mass cores for the expected massive star formation (assuming a core-mass-to-stellar-mass conversion of 30$\%$) in this region implies further core mass growth \citep[also see][]{Morii_2023}. There are other models such as global hierarchical collapse model and inertial-inflow model \citep{Vazquez-Semadeni_2019, Padoan_2020}, but more comprehensive scenario is still needed to better explain the observation results \citep{Li_2023}. 

\subsection{Protostellar Activities and Star Formation Signatures in the Cloud}

Molecular line emissions of the CO 2-1 and SiO 5-4 transitions are detected from some of the dense cores. The CO molecular line emission serves as a typical tracer for protostellar outflows due to its high abundance ($\sim$10$^{-4}$ relative to hydrogen) compared with other molecular tracers. Molecules such as SiO and CH$_3$OH may be depleted in cold and dense regions, with a relative abundance of approximately 10$^{-12}-$10$^{-10}$ \citep{Zhang_2015}. However, in protostellar outflows, their abundances increase significantly due to shocks originating from interactions between the protostellar wind and the surrounding medium, facilitating the release of Si and CH$_3$OH from the dust grain \citep{Zhang_2015}. Therefore, protostellar outflows traced by SiO and CH$_3$OH serve as a cross-reference for outflow detection. The widely traced outflows shown in Figures \ref{fig:n_sig_outflow} and \ref{fig:mosaic_outflow} reveal ongoing active protostellar activities in the embedded core of the G14.225 cloud. Outflows with different orientations that originate from the common dense core regions, for example, the northern pointing of the mosaic north field, may indicate multiple protostars forming in the areas. Cores that have no or faint molecular line emission from the CO 2-1, SiO 5-4, and CH$_3$OH 4-3 may harbor no protostars and are prestellar candidates. 

The most massive outflows in the observed region have a dynamical time scale of $0.4-3.4 \times 10^4$ years, an outflow mass rate of $0.6-31.2 \times 10^{-6}$ M$_{\odot}$ yr$^{-1}$, a mechanical force of $0.03-3.69 \times10^{-4}$  M$_{\odot}$ km s$^{-1}$ yr$^{-1}$, and an outflow luminosity of $0.06-21.45 \times 10^{32}$ erg s$^{-1}$. The calculation uses the primary beam corrected CO maps, and does not consider the inclination angle of the outflow axis with respect to the line of sight direction. We compare outflow parameters with other interferometry observations, since observations from single-dish telescopes generally report larger outflow masses \citep[e.g.][]{Zhang_2005}, but may not spatially resolve multiple flows. Compared with the ALMA survey of massive IRDCs by \cite{Li_2020b}, we found that the outflow parameters in this study are consistent within one order of magnitude. The consistency implies that the protostellar outflows in the G14.225 cloud share similar dynamical properties as those in other massive infrared dark clumps, and are still in an early evolutionary stage of star and cluster formation. Moreover, compared to the outflows in more evolved protoclusters studied by \cite{Baug_2021}, outflow parameters in this study have smaller outflow rates, mechanical force and luminosities by one to two orders of magnitude. Since outflow rates correlate with rates of mass accretion onto protostars, this implies lower accretion rates among protostars in the G14.225 cloud as compared to those in \cite{Baug_2021}. This increasing trend in accretion rates over evolutionary stage of massive protostars, as proposed by \citet{Zhang_2015}, has also been reported in \citet{Li_2020b}.

\begin{figure*}[h!]
\centering
\includegraphics[width=15.0cm]{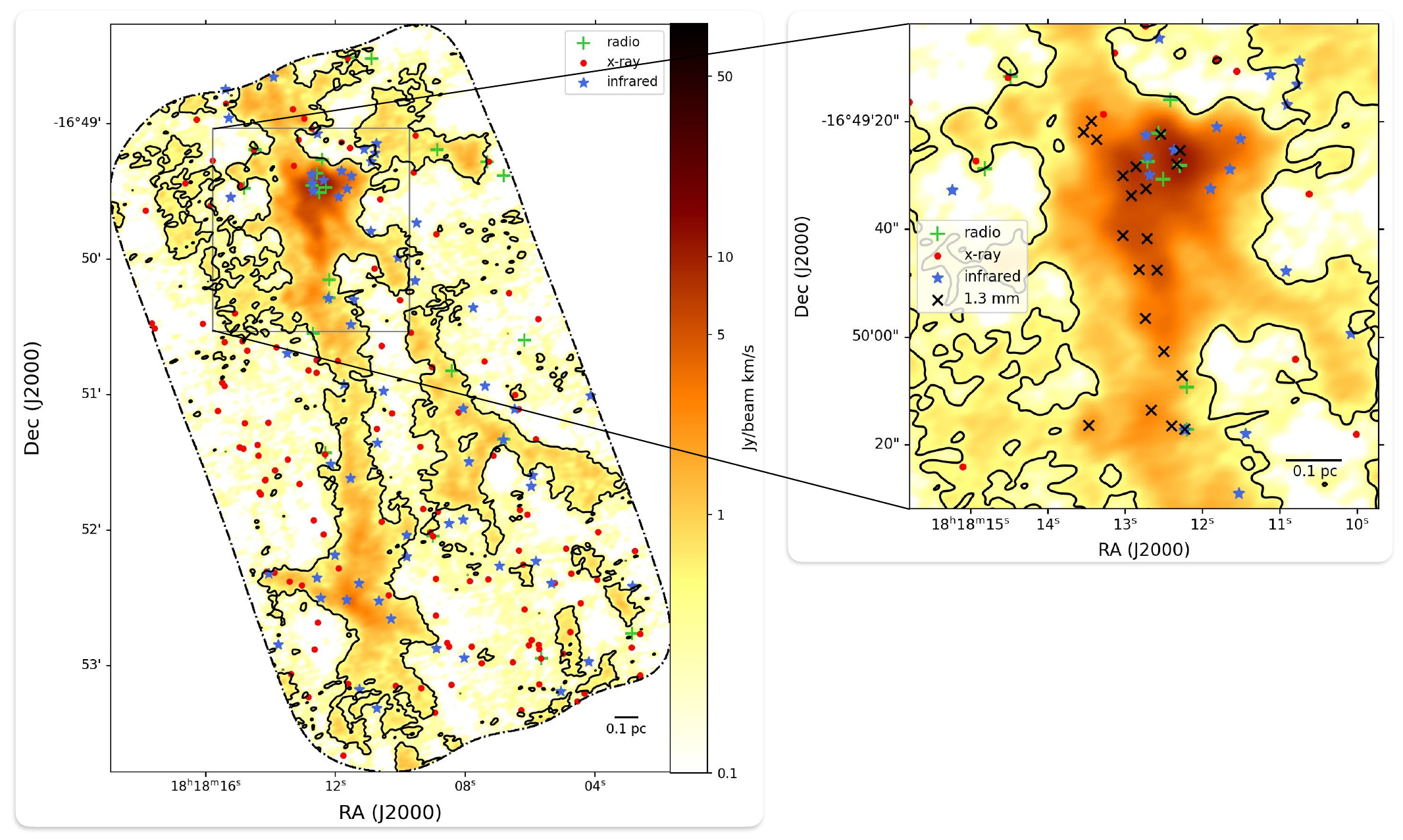}
\includegraphics[width=15.0cm]{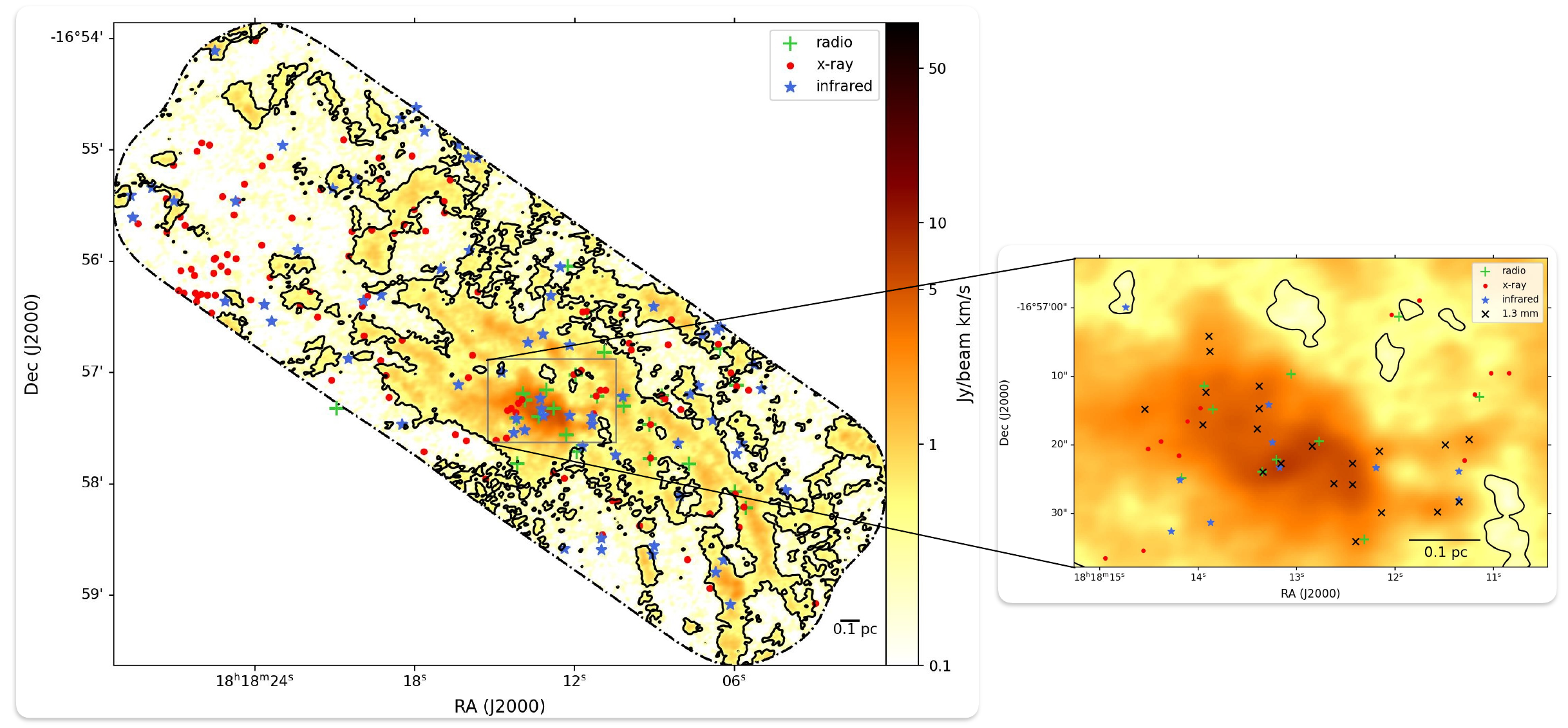}
\caption{Larger-scale view of the G14N and G14S hub-filament region with the dense cores identified in this study shown in the zoom-in view. Infrared (blue star), X-ray (red dot), and radio (green plus sign) sources from studies of \cite{Povich_2010,Povich_2016,Elena_2024} overlaid on the N$_2$H$^+$ (1-0) integrated emission map from \citep{Chen_2019}. The map is colored by the integrated intensity level of N$_2$H$^+$ (1-0) in the unit of Jy~beam$^{-1}$~km~s$^{-1}$. Black contour denotes the emission level of 3$\sigma$, where $\sigma$ is the rms noise level of the N$_2$H$^+$ (1-0) integrated emission. Sources located within the 3$\sigma$ contour are regarded as inside the dense "cloud" region, with statistics summarized in Table~\ref{tab:diff_wvln}. The right zoom-in plots show the detailed distribution pattern of these sources and the 1.3 mm source (black cross) analyzed in this study in the G14N mosaic and G14S mosaic regions.}
\label{fig:multi-wavelength}
\end{figure*}

In addition to protostellar outflows in the dense regions of molecular clouds that trace the deeply embedded phase of star formation, young stellar objects (YSO) are also probed by infrared, X-ray, and radio surveys of G14.225 \citep{Povich_2010, Povich_2016, Elena_2024}. Figure \ref{fig:multi-wavelength} presents an overlay of these sources on the N$_2$H$^+$ (1-0) emission \citep{Chen_2019}. The statistics of these observations are tabulated in Table \ref{tab:diff_wvln}, classified by the emission of the N$_2$H$^+$ (1-0) from the ALMA observations, where an emission level $>$3$\sigma$ ($\sigma$ is the noise level of the integrated N$_2$H$^+$ map) is considered as inside of the dense cloud. From the radio continuum survey performed by \cite{Elena_2024} using the Karl G. Jansky Very Large Array (VLA), properties of the radio continuum emission in the C-band (4–8 GHz, $\sim$6 cm) and X-band (8–12 GHz, $\sim$3.6 cm) are analyzed. The spectral indices of the emission reveal a population of objects with thermal and non-thermal emission. The non-thermal emission usually results from the gyrosynchrotron radiation, which is likely linked to surface magnetic activities of young stars, and is one of the characteristics of Class II/III YSOs \citep{Elena_2024}. Sources with thermal emission, on the other hand, are typically due to Bremsstrahlung radiation when free electrons accelerate in the electric field of H$^-$, and can probe jets powered by YSOs. A larger faction of radio sources are detected inside the cloud compared to the outside areas. Radio sources in the northern regions exhibit dominantly non-thermal properties while the radio sources in the southern fields tend to be more thermal. This is consistent with the fact that the southern cloud is at an earlier evolutionary stage than that of the northern cloud.

Infrared and X-ray sources are summarized from the Mid-IR Excess Source (MIRES) Catalog and the Chandra X-Ray catalog in \cite{Povich_2016}, respectively. X-ray emission is typically associated with magnetic reconnections in the chromosphere and corona of pre-main sequence stars. For the infrared sources, their evolution stages are inferred from the spectral energy distributions obtained with the Spitzer Space Telescope. Significant fractions of YSO sources are observed outside the dense "cloud" regions, implying that there may be a considerable amount of young stars already formed in the diffuse gas regions.

YSOs, and radio continuum and X-ray sources are found to be associated with both hubs in G14.225 N and G14.225 S where dense cores and protostellar outflows are identified in the ALMA observations. However, these sources do not coincide with the peak of dense cores {\fontfamily{qcr}\selectfont leaf} identified from the continuum observations at 1.3 mm, since only 7 pairs of sources are spatially separated within 1.5$''$ criteria which corresponds to the synthesized beam size of this study. These YSOs identified from infrared, X-ray and radio bands likely represent a populations of young stars at an more evolved stage than the embedded protostellar populations powering protostellar outflows. \cite{Povich_2016} found that the population of YSOs revealed in the IR observations has a maximum mass of less than 10 M$_\odot$, and there are no O-type stars in this cloud. The more evolved population of stars has likely completed the active accretion phase, and gained the majority of stellar masses. The deeply embedded population of protostars are currently accreting as shown by powerful outflows. Some of them may become stars of $>$ 10 M$_\odot$ when the accretion is complete. 
This progression in evolutionary stages of stellar populations indicates that most massive stars experience longer accretion than lower mass stars in a cluster.

\begin{table*}[h]
\caption{Comparison of Radio, Infrared, and X-ray Sources}
\centering
\label{tab:diff_wvln}

\begin{tabular}{cccccc}
\hline
\multicolumn{1}{|l}{\textbf{(a) Radio sources}} & & & & & \multicolumn{1}{c|}{}\\ \hline
\multicolumn{1}{|c}{Northern region}                      & \multicolumn{1}{c}{(22 sources)}                     &                                    & \multicolumn{1}{|c}{Southern region}                  & \multicolumn{1}{c}{(25 sources)}                        &    \multicolumn{1}{c|}{}                                \\ \hline
\hline
\multicolumn{1}{|c|}{radio property} & \multicolumn{1}{c|}{inside cloud} & \multicolumn{1}{c|}{outside cloud} & \multicolumn{1}{c|}{radio property} & \multicolumn{1}{c|}{inside cloud} & \multicolumn{1}{c|}{outside cloud} \\ \hline
\multicolumn{1}{|c|}{thermal}        & \multicolumn{1}{c|}{0}            & \multicolumn{1}{c|}{0}             & \multicolumn{1}{c|}{thermal}        & \multicolumn{1}{c|}{8}            & \multicolumn{1}{c|}{1}             \\ \hline
\multicolumn{1}{|c|}{flat}           & \multicolumn{1}{c|}{2}            & \multicolumn{1}{c|}{0}             & \multicolumn{1}{c|}{flat}           & \multicolumn{1}{c|}{0}            & \multicolumn{1}{c|}{1}             \\ \hline
\multicolumn{1}{|c|}{non-thermal}    & \multicolumn{1}{c|}{8}            & \multicolumn{1}{c|}{2}             & \multicolumn{1}{c|}{non-thermal}    & \multicolumn{1}{c|}{6}            & \multicolumn{1}{c|}{0}             \\ \hline
\multicolumn{1}{|c|}{variable}       & \multicolumn{1}{c|}{0}            & \multicolumn{1}{c|}{3}             & \multicolumn{1}{c|}{variable}       & \multicolumn{1}{c|}{4}            & \multicolumn{1}{c|}{2}             \\ \hline
\multicolumn{1}{|c|}{unclassified}   & \multicolumn{1}{c|}{3}            & \multicolumn{1}{c|}{4}             & \multicolumn{1}{c|}{unclassified}   & \multicolumn{1}{c|}{2}            & \multicolumn{1}{c|}{1}             \\ \hline
\hline
\multicolumn{1}{|c|}{total}          & \multicolumn{1}{c|}{13}           & \multicolumn{1}{c|}{9}             & \multicolumn{1}{c|}{total}          & \multicolumn{1}{c|}{20}           & \multicolumn{1}{c|}{5}             \\ \hline
\\
\hline
\multicolumn{1}{|l}{\textbf{(b) IR sources}} & & & & & \multicolumn{1}{c|}{}\\ \hline
\multicolumn{1}{|c}{Northern region}  &  (62 sources)                 &                          & \multicolumn{1}{|c}{Southern region} &   (66 sources)    &  \multicolumn{1}{c|}{}       \\ 
\hline
\hline
\multicolumn{1}{|c|}{YSO stage}    & \multicolumn{1}{c|}{inside cloud} & \multicolumn{1}{c|}{outside cloud} & \multicolumn{1}{c|}{YSO stage}    & \multicolumn{1}{c|}{inside cloud} & \multicolumn{1}{c|}{outside cloud} \\ \hline
\multicolumn{1}{|c|}{stage 0/I}    & \multicolumn{1}{c|}{16}                              & \multicolumn{1}{c|}{13}                            & \multicolumn{1}{c|}{stage 0/I}    & \multicolumn{1}{c|}{14}                              & \multicolumn{1}{c|}{9}                             \\ \hline
\multicolumn{1}{|c|}{stage II/III} & \multicolumn{1}{c|}{11}                              & \multicolumn{1}{c|}{17}                            & \multicolumn{1}{c|}{stage II/III} & \multicolumn{1}{c|}{14}                              & \multicolumn{1}{c|}{16}                            \\ \hline
\multicolumn{1}{|c|}{ambiguous}    & \multicolumn{1}{c|}{3}                               & \multicolumn{1}{c|}{2}                             & \multicolumn{1}{c|}{ambiguous}    & \multicolumn{1}{c|}{9}                               & \multicolumn{1}{c|}{4}                             \\ \hline
\hline
\multicolumn{1}{|c|}{total}        & \multicolumn{1}{c|}{30}                              & \multicolumn{1}{c|}{32}                            & \multicolumn{1}{c|}{total}        & \multicolumn{1}{c|}{37}                              & \multicolumn{1}{c|}{29}                            \\ \hline
\\
\hline
\multicolumn{1}{|l}{\textbf{(c) X-ray sources}} & & & & & \multicolumn{1}{c|}{}\\ \hline
\multicolumn{1}{|c}{Northern region}   & (146 sources, & 106 PCMs \footnote{PCM: Probable Complex Member. The criteria of PCM classification include MIR excess emission, X-ray variability, X-ray median energy and IR extinction, proximity to other PCM and manual selection \citep{Povich_2016}. Other detected sources are either unclassified or classified to background and foreground and thus excluded from the statistics.})  & \multicolumn{1}{|c}{Southern region} & (133 sources,  &  \multicolumn{1}{c|}{80 PCMs)}  \\
\hline
\hline
\multicolumn{1}{|c|}{YSO stage}    & \multicolumn{1}{c|}{inside cloud} & \multicolumn{1}{c|}{outside cloud} & \multicolumn{1}{c|}{YSO stage}    & \multicolumn{1}{c|}{inside cloud} & \multicolumn{1}{c|}{outside cloud} \\ \hline
\multicolumn{1}{|c|}{stage 0/I}    & \multicolumn{1}{c|}{5}                               & \multicolumn{1}{c|}{6}                             & \multicolumn{1}{c|}{stage 0/I}    & \multicolumn{1}{c|}{1}                               & \multicolumn{1}{c|}{3}                             \\ \hline
\multicolumn{1}{|c|}{stage II/III} & \multicolumn{1}{c|}{4}                               & \multicolumn{1}{c|}{13}                            & \multicolumn{1}{c|}{stage II/III} & \multicolumn{1}{c|}{1}                               & \multicolumn{1}{c|}{8}                             \\ \hline
\multicolumn{1}{|c|}{ambiguous}    & \multicolumn{1}{c|}{0}                               & \multicolumn{1}{c|}{1}                             & \multicolumn{1}{c|}{ambiguous}    & \multicolumn{1}{c|}{3}                               & \multicolumn{1}{c|}{6}                             \\ \hline
\multicolumn{1}{|c|}{unclassified} & \multicolumn{1}{c|}{13}                              & \multicolumn{1}{c|}{64}                            & \multicolumn{1}{c|}{unclassified} & \multicolumn{1}{c|}{14}                              & \multicolumn{1}{c|}{44}                            \\ \hline \hline
\multicolumn{1}{|c|}{total}        & \multicolumn{1}{c|}{22}                              & \multicolumn{1}{c|}{84}                            & \multicolumn{1}{c|}{total}        & \multicolumn{1}{c|}{19}                              & \multicolumn{1}{c|}{61}                            \\ \hline
\end{tabular}

\end{table*}

\section{Conclusion} \label{sec:conclusion}

We performed an analysis of dense cores and protostellar outflows in the hub-filament system of IRDC G14.225-0.506 using ALMA band 6 observations in the 1.3 mm continuum emission and the CO J=2-1 and SiO J=5-4 transitions. The main results are summarized as follows.\\

(1) We detected 221 dense core structures in the 1.3 mm continuum emission. The temperatures of these structures are $\sim$15$-$22 K, and the densities are 10$^{22}-$10$^{23}$ cm$^{-2}$. The most massive leaf structure has a mass of 20.4 $M_\odot$.\\

(2) The dense core structures are dominated by transonic and supersonic motions. Virial parameters of the cores decrease with an increasing density, implying that the effect of gravity becomes more significant than thermal and turbulent support in higher density regions.\\

(3) Molecular outflows are widely detected in some of the dense cores using the CO 2-1 and SiO 5-4 emission, indicating strong protostellar activities in the cloud.\\

(4) YSOs, radio continuum and X-ray sources towards G14.225 from previous studies suggest a distributed stellar population in the region. The majority of these sources do not coincide with the emission peaks detected in the 1.3 mm continuum, and they represent some more evolved young stars that have accreted most of their mass, while the embedded protostars are still active accreting as indicated by outflows.\\

%% Also note that the akcnowlodgment environment does not support long amounts of text. If you have a lot of people and institutions to acknowledge, do not use this command. Instead, create a new \section{Acknowledgments}.
\begin{acknowledgments}

We thank the anonymous referee for constructive comments that helped improve the presentation and clarity of the paper. Y.Z. acknowledges support from the CfA summer internship and SURE program offered by Department of Physics at The Chinese University of Hong Kong. X.P. is supported by the Smithsonian Astrophysical Observatory (SAO) Predoctoral Fellowship Program.
This paper makes use of the following ALMA data: ADS/JAO.ALMA$\#$2017.1.00793.S. ALMA is a partnership of ESO (representing its member states), NSF (USA) and NINS (Japan), together with NRC (Canada), MOST and ASIAA (Taiwan), and KASI (Republic of Korea), in cooperation with the Republic of Chile. The Joint ALMA Observatory is operated by ESO, AUI/NRAO and NAOJ. This research made use of astrodendro, a Python package to compute dendrograms of Astronomical data (http://www.dendrograms.org/). 

\end{acknowledgments}


\vspace{5mm}
\facilities{ALMA}


\software{Astropy \citep{astropy:2022, astropy:2018, astropy:2013}, 
 AstroDendro \citep{astrodendro_2019}, CASA \citep{CASA_Team_2022}, Matplotlib \citep{Hunter_2007}, SAOImageDS9 \citep{SAO_2000}}



\bibliography{sample631_arxiv_version}{}
\bibliographystyle{aasjournal}

\appendix

%\setcounter{table}{2}
%\begin{center}
\begin{longtable*}[ht!]{cccccccccccc}
\caption{Physical parameters of core structures} 
\label{tab:continuum_phy_para} \\
\hline \hline 
\multicolumn{1}{c}{$id$} & 
\multicolumn{1}{c}{$R.A.$} & 
\multicolumn{1}{c}{$Decl.$}& \multicolumn{1}{c}{$size$} & \multicolumn{1}{c}{$PA$} & \multicolumn{1}{c}{$r$} & \multicolumn{1}{c}{$flux$} & \multicolumn{1}{c}{$T_d$} & \multicolumn{1}{c}{$Mass$} & \multicolumn{1}{c}{$density$} \\ 
 & $(hh:mm:ss)$ & $(dd:mm:ss)$ & $(\ '' \times \ '')$ & $(^{\circ})$ &  (pc) &  $(mJy)$ & $(K)$ & $(M_{\odot})$ & $(cm^{-2})$ \\
\endfirsthead

\multicolumn{10}{c}%
{{ \tablename\ \thetable{} -- Physical parameters of core structures (continued)}} \\
\hline \hline \multicolumn{1}{c}{$id$} & \multicolumn{1}{c}{$R.A.$} & \multicolumn{1}{c}{$Decl.$}& \multicolumn{1}{c}{$size$} & \multicolumn{1}{c}{$PA$} & \multicolumn{1}{c}{$r$} & \multicolumn{1}{c}{$flux$} & \multicolumn{1}{c}{$T_d$} & \multicolumn{1}{c}{$Mass$} & \multicolumn{1}{c}{$density$}  \\ 
 & $(hh:mm:ss)$ & $(dd:mm:ss)$ & $(\ '' \times \ '')$ & $(^{\circ})$ &  (pc) &  $(mJy)$ & $(K)$ & $(M_{\odot})$ & $(cm^{-2})$ \\
 \hline
\endhead

\hline 
\multicolumn{10}{r}{$id$ with * is {\fontfamily{qcr}\selectfont leaf} structure. $size$ with * appears to be point sources and can not be deconvolved.}\\
\multicolumn{10}{r}{$flux$ is computed with primary beam correction. $T_d$ is obtained from \cite{Busquet_2013}.}\\
\multicolumn{10}{r}{{Continued on next page}}  \\
\endfoot
\endlastfoot

\hline
N.4$-$0 & 18h18m11.38s & -16d52m33.48s & 10.24 $ \times $ 3.96 & 46.17 & 0.061 & 177.27 & 22.24 & 12.78 & 4.85e+22 \\
N.4$-$1 & 18h18m11.38s & -16d52m33.46s & 9.91 $ \times $ 3.72 & 45.53 & 0.058 & 172.04 & 22.50 & 12.22 & 5.10e+22 \\
N.4$-$2 & 18h18m11.41s & -16d52m33.54s & 7.72 $ \times $ 2.7 & 44.04 & 0.044 & 137.80 & 24.06 & 9.00 & 6.65e+22 \\
N.4$-$3$^*$ & 18h18m10.94s & -16d52m40.30s & 1.27 $ \times $ 0.9$^*$ & -43.92 & 0.010 & 2.50 & 20.24 & 0.20 & 2.72e+22 \\
N.4$-$4$^*$ & 18h18m10.67s & -16d52m39.24s & 1.45 $ \times $ 0.54$^*$ & -38.70 & 0.008 & 0.89 & 18.37 & 0.08 & 1.64e+22 \\
N.4$-$5 & 18h18m11.42s & -16d52m33.42s & 7.08 $ \times $ 2.67 & 43.58 & 0.042 & 132.80 & 24.29 & 8.57 & 6.98e+22 \\
N.4$-$6$^*$ & 18h18m10.95s & -16d52m38.10s & 1.71 $ \times $ 0.57$^*$ & -42.59 & 0.009 & 1.69 & 19.65 & 0.14 & 2.25e+22 \\
N.4$-$7 & 18h18m11.43s & -16d52m33.36s & 6.45 $ \times $ 2.52 & 40.88 & 0.039 & 126.01 & 24.50 & 8.05 & 7.61e+22 \\
N.4$-$8 & 18h18m11.45s & -16d52m33.25s & 5.62 $ \times $ 1.55 & 29.88 & 0.028 & 112.79 & 26.38 & 6.58 & 1.16e+23 \\
N.4$-$9$^*$ & 18h18m11.27s & -16d52m37.28s & 1.32 $ \times $ 0.99$^*$ & 113.04 & 0.011 & 16.79 & 21.59 & 1.26 & 1.48e+23 \\
N.4$-$10$^*$ & 18h18m10.44s & -16d52m37.40s & 1.58 $ \times $ 0.94$^*$ & 88.31 & 0.012 & 3.91 & 17.31 & 0.39 & 4.02e+22 \\
N.4$-$11 & 18h18m11.44s & -16d52m33.31s & 5.87 $ \times $ 1.95 & 34.34 & 0.032 & 117.06 & 25.11 & 7.25 & 9.76e+22 \\
N.4$-$12$^*$ & 18h18m11.49s & -16d52m32.54s & 2.44 $ \times $ 1.14 & 13.35 & 0.016 & 85.32 & 27.75 & 4.68 & 2.58e+23 \\
N.4$-$13$^*$ & 18h18m11.09s & -16d52m35.74s & 1.18 $ \times $ 0.95$^*$ & 98.93 & 0.010 & 2.71 & 19.34 & 0.23 & 3.22e+22 \\
N.4$-$14$^*$ & 18h18m10.91s & -16d52m36.17s & 1.03 $ \times $ 0.53$^*$ & 99.30 & 0.007 & 1.28 & 17.20 & 0.13 & 3.65e+22 \\
N.4$-$15$^*$ & 18h18m10.67s & -16d52m31.18s & 1.07 $ \times $ 0.82$^*$ & 108.93 & 0.009 & 1.26 & 17.09 & 0.13 & 2.25e+22 \\
N.4$-$16$^*$ & 18h18m12.07s & -16d52m29.00s & 1.88 $ \times $ 0.91$^*$ & 67.89 & 0.013 & 1.37 & 29.69 & 0.07 & 6.28e+21 \\
N.4$-$17$^*$ & 18h18m11.69s & -16d52m26.77s & 1.12 $ \times $ 0.6$^*$ & 71.68 & 0.008 & 1.03 & 23.07 & 0.07 & 1.62e+22 \\
\hline
\hline
N.5$-$0 & 18h18m06.87s & -16d51m21.42s & 15.42 $ \times $ 9.28 & 78.73 & 0.115 & 61.80 & 14.56 & 7.83 & 8.43e+21 \\
N.5$-$1 & 18h18m07.51s & -16d51m27.70s & 5.96 $ \times $ 1.09 & -14.35 & 0.024 & 10.11 & 14.08 & 1.34 & 3.18e+22 \\
N.5$-$2$^*$ & 18h18m07.57s & -16d51m30.90s & 1.22 $ \times $ 0.65$^*$ & -25.63 & 0.009 & 1.25 & 14.44 & 0.16 & 3.13e+22 \\
N.5$-$3$^*$ & 18h18m07.49s & -16d51m26.48s & 3.13 $ \times $ 1.12$^*$ & -7.49 & 0.018 & 2.86 & 14.13 & 0.38 & 1.66e+22 \\
N.5$-$4 & 18h18m06.78s & -16d51m20.33s & 14.75 $ \times $ 3.0 & 61.22 & 0.064 & 44.27 & 15.17 & 5.29 & 1.84e+22 \\
N.5$-$5$^*$ & 18h18m06.00s & -16d51m26.38s & 1.45 $ \times $ 0.75 & 84.28 & 0.010 & 8.77 & 14.56 & 1.11 & 1.57e+23 \\
N.5$-$6 & 18h18m06.91s & -16d51m19.22s & 8.56 $ \times $ 2.72 & 59.33 & 0.046 & 29.35 & 15.30 & 3.47 & 2.30e+22 \\
N.5$-$7$^*$ & 18h18m06.40s & -16d51m23.93s & 3.23 $ \times $ 0.75$^*$ & 67.93 & 0.015 & 1.71 & 14.56 & 0.22 & 1.37e+22 \\
N.5$-$8 & 18h18m06.95s & -16d51m18.87s & 6.46 $ \times $ 2.7 & 60.87 & 0.040 & 25.93 & 15.30 & 3.07 & 2.71e+22 \\
N.5$-$9$^*$ & 18h18m06.91s & -16d51m19.27s & 4.79 $ \times $ 2.46 & 75.82 & 0.033 & 22.10 & 14.63 & 2.78 & 3.63e+22 \\
N.5$-$10$^*$ & 18h18m07.25s & -16d51m15.19s & 1.19 $ \times $ 0.45 & -24.93 & 0.007 & 2.94 & 15.02 & 0.36 & 1.03e+23 \\
N.5$-$11$^*$ & 18h18m08.02s & -16d51m15.02s & 1.67 $ \times $ 0.96$^*$ & 95.54 & 0.012 & 3.37 & 14.56 & 0.43 & 4.08e+22 \\
N.5$-$12$^*$ & 18h18m08.03s & -16d51m06.42s & 2.06 $ \times $ 1.22$^*$ & 55.71 & 0.015 & 12.53 & 14.56 & 1.59 & 9.72e+22 \\
N.5$-$13$^*$ & 18h18m06.48s & -16d51m06.37s & 1.28 $ \times $ 0.83$^*$ & -31.87 & 0.010 & 0.97 & 14.56 & 0.12 & 1.79e+22 \\
\hline
\hline
N.7$-$0 & 18h18m11.59s & -16d51m35.41s & 14.16 $ \times $ 3.16 & 1.31 & 0.064 & 59.57 & 14.56 & 7.55 & 2.60e+22 \\
N.7$-$1$^*$ & 18h18m11.03s & -16d51m46.99s & 2.37 $ \times $ 0.33 & 53.00 & 0.008 & 1.81 & 14.56 & 0.23 & 4.56e+22 \\
N.7$-$2$^*$ & 18h18m11.63s & -16d51m44.11s & 2.48 $ \times $ 1.06 & -16.12 & 0.016 & 3.76 & 14.56 & 0.48 & 2.80e+22 \\
N.7$-$3 & 18h18m11.59s & -16d51m34.62s & 9.64 $ \times $ 2.76 & 3.31 & 0.049 & 40.08 & 14.56 & 5.08 & 2.94e+22 \\
N.7$-$4$^*$ & 18h18m11.57s & -16d51m36.81s & 2.85 $ \times $ 2.12 & 69.88 & 0.024 & 23.08 & 14.56 & 2.92 & 7.47e+22 \\
N.7$-$5 & 18h18m11.63s & -16d51m29.52s & 5.85 $ \times $ 0.22 & -6.45 & 0.011 & 8.01 & 14.56 & 1.01 & 1.21e+23 \\
N.7$-$6$^*$ & 18h18m11.64s & -16d51m30.27s & 3.99 $ \times $ 1.32$^*$ & -9.03 & 0.022 & 5.80 & 14.56 & 0.73 & 2.15e+22 \\
N.7$-$7$^*$ & 18h18m11.61s & -16d51m25.03s & 1.22 $ \times $ 0.8$^*$ & 13.03 & 0.009 & 1.23 & 14.56 & 0.16 & 2.46e+22 \\
\hline
\hline
N$-$0$^*$ & 18h18m12.64s & -16d50m25.21s & 0.98 $ \times $ 0.88$^*$ & -26.91 & 0.009 & 1.80 & 19.34 & 0.15 & 2.75e+22 \\
N$-$1 & 18h18m12.50s & -16d49m34.55s & 30.0 $ \times $ 9.06 & -5.92 & 0.158 & 1232.01 & 18.57 & 112.02 & 6.34e+22 \\
N$-$2 & 18h18m12.61s & -16d50m14.18s & 6.38 $ \times $ 4.48 & 54.06 & 0.051 & 53.13 & 14.71 & 6.64 & 3.58e+22 \\
N$-$3$^*$ & 18h18m13.47s & -16d50m16.37s & 1.8 $ \times $ 1.22 & -29.31 & 0.014 & 8.81 & 15.14 & 1.06 & 7.40e+22 \\
N$-$4$^*$ & 18h18m12.66s & -16d50m13.56s & 3.33 $ \times $ 2.22 & -18.92 & 0.026 & 34.39 & 14.81 & 4.26 & 8.87e+22 \\
N$-$5 & 18h18m12.29s & -16d50m16.84s & 2.86 $ \times $ 0.04 & 78.39 & 0.003 & 8.80 & 16.78 & 0.92 & 1.12e+24 \\
N$-$6$^*$ & 18h18m12.23s & -16d50m17.05s & 1.36 $ \times $ 1.02$^*$ & 82.93 & 0.011 & 4.98 & 16.83 & 0.52 & 5.72e+22 \\
N$-$7$^*$ & 18h18m12.40s & -16d50m16.54s & 1.12 $ \times $ 0.79$^*$ & 119.36 & 0.009 & 1.78 & 16.78 & 0.19 & 3.25e+22 \\
N$-$8 & 18h18m12.61s & -16d50m14.13s & 6.4 $ \times $ 4.58 & 51.64 & 0.052 & 53.55 & 14.60 & 6.76 & 3.55e+22 \\
N$-$9 & 18h18m12.49s & -16d49m32.95s & 23.97 $ \times $ 9.02 & -8.57 & 0.141 & 1175.69 & 19.10 & 103.01 & 7.33e+22 \\
N$-$10 & 18h18m12.49s & -16d49m32.83s & 23.72 $ \times $ 8.92 & -8.94 & 0.140 & 1159.48 & 19.24 & 100.61 & 7.33e+22 \\
N$-$12 & 18h18m12.49s & -16d49m32.72s & 23.47 $ \times $ 8.86 & -9.23 & 0.138 & 1147.09 & 19.38 & 98.61 & 7.30e+22 \\
N$-$13 & 18h18m12.57s & -16d50m00.48s & 9.23 $ \times $ 3.34 & 25.11 & 0.053 & 108.87 & 15.82 & 12.29 & 6.13e+22 \\
N$-$14$^*$ & 18h18m12.26s & -16d50m07.19s & 1.64 $ \times $ 0.96$^*$ & -33.32 & 0.012 & 1.64 & 16.83 & 0.17 & 1.67e+22 \\
N$-$15 & 18h18m12.58s & -16d50m00.34s & 8.79 $ \times $ 3.23 & 24.17 & 0.051 & 103.79 & 15.79 & 11.75 & 6.36e+22 \\
N$-$16$^*$ & 18h18m12.50s & -16d50m02.65s & 4.18 $ \times $ 1.6 & 1.06 & 0.025 & 45.21 & 15.27 & 5.36 & 1.23e+23 \\
N$-$18$^*$ & 18h18m12.74s & -16d49m56.51s & 2.35 $ \times $ 1.47 & 47.88 & 0.018 & 20.62 & 16.49 & 2.20 & 9.81e+22 \\
N$-$20 & 18h18m12.49s & -16d49m30.43s & 15.18 $ \times $ 7.79 & -26.83 & 0.104 & 1031.30 & 20.31 & 83.46 & 1.09e+23 \\
N$-$21 & 18h18m12.49s & -16d49m30.39s & 15.13 $ \times $ 7.64 & -27.31 & 0.103 & 1022.83 & 20.32 & 82.73 & 1.10e+23 \\
N$-$22$^*$ & 18h18m13.19s & -16d49m49.62s & 2.42 $ \times $ 0.52 & 46.44 & 0.011 & 3.44 & 22.60 & 0.24 & 2.97e+22 \\
N$-$23 & 18h18m12.48s & -16d49m29.86s & 13.91 $ \times $ 6.75 & -32.79 & 0.093 & 939.57 & 20.56 & 74.82 & 1.23e+23 \\
N$-$24$^*$ & 18h18m12.59s & -16d49m47.56s & 2.63 $ \times $ 1.21$^*$ & 42.11 & 0.017 & 8.17 & 19.27 & 0.71 & 3.43e+22 \\
N$-$26$^*$ & 18h18m12.82s & -16d49m47.46s & 1.87 $ \times $ 0.77$^*$ & 48.63 & 0.011 & 4.79 & 19.47 & 0.41 & 4.39e+22 \\
N$-$27 & 18h18m13.71s & -16d49m41.03s & 6.21 $ \times $ 1.29 & -3.54 & 0.027 & 10.74 & 20.56 & 0.86 & 1.64e+22 \\
N$-$28 & 18h18m12.48s & -16d49m29.72s & 13.54 $ \times $ 6.51 & -35.20 & 0.090 & 925.88 & 20.59 & 73.59 & 1.28e+23 \\
N$-$29$^*$ & 18h18m12.58s & -16d49m44.99s & 0.82 $ \times $ 0.73$^*$ & 54.57 & 0.007 & 0.70 & 20.41 & 0.06 & 1.46e+22 \\
N$-$30 & 18h18m12.98s & -16d49m41.21s & 4.0 $ \times $ 2.49 & 71.70 & 0.030 & 99.34 & 19.34 & 8.56 & 1.32e+23 \\
N$-$31$^*$ & 18h18m11.73s & -16d49m44.43s & 1.29 $ \times $ 0.53$^*$ & 116.31 & 0.008 & 1.78 & 18.15 & 0.17 & 3.75e+22 \\
N$-$32 & 18h18m13.72s & -16d49m41.33s & 4.91 $ \times $ 1.17 & -5.73 & 0.023 & 8.75 & 20.56 & 0.70 & 1.87e+22 \\
N$-$33 & 18h18m13.72s & -16d49m42.27s & 2.78 $ \times $ 1.24 & 2.79 & 0.018 & 6.79 & 21.11 & 0.52 & 2.33e+22 \\
N$-$34$^*$ & 18h18m13.72s & -16d49m43.08s & 1.52 $ \times $ 1.28$^*$ & -22.51 & 0.013 & 3.54 & 21.78 & 0.26 & 2.08e+22 \\
N$-$35$^*$ & 18h18m11.86s & -16d49m42.73s & 1.49 $ \times $ 0.96$^*$ & 124.35 & 0.011 & 2.83 & 18.63 & 0.26 & 2.77e+22 \\
N$-$36$^*$ & 18h18m13.03s & -16d49m41.13s & 1.85 $ \times $ 1.49 & 84.50 & 0.016 & 72.05 & 19.16 & 6.29 & 3.52e+23 \\
N$-$37 & 18h18m11.45s & -16d49m39.54s & 3.9 $ \times $ 2.05 & -5.31 & 0.027 & 23.93 & 18.04 & 2.26 & 4.36e+22 \\
N$-$38 & 18h18m11.45s & -16d49m39.26s & 4.65 $ \times $ 1.99 & -3.30 & 0.029 & 25.44 & 18.04 & 2.40 & 3.99e+22 \\
N$-$39 & 18h18m11.42s & -16d49m40.39s & 3.32 $ \times $ 0.84 & -32.18 & 0.016 & 13.69 & 18.05 & 1.29 & 7.14e+22 \\
N$-$40$^*$ & 18h18m11.42s & -16d49m42.12s & 1.42 $ \times $ 0.44$^*$ & 134.19 & 0.008 & 2.37 & 18.13 & 0.22 & 5.42e+22 \\
N$-$41$^*$ & 18h18m12.72s & -16d49m41.71s & 1.32 $ \times $ 1.12$^*$ & -20.17 & 0.012 & 6.03 & 19.87 & 0.50 & 5.20e+22 \\
N$-$42 & 18h18m11.42s & -16d49m39.97s & 3.34 $ \times $ 0.85$^*$ & -39.58 & 0.016 & 9.05 & 18.00 & 0.86 & 4.63e+22 \\
N$-$43$^*$ & 18h18m11.46s & -16d49m40.71s & 1.46 $ \times $ 0.59$^*$ & -35.82 & 0.009 & 3.86 & 18.05 & 0.36 & 6.57e+22 \\
N$-$44$^*$ & 18h18m13.71s & -16d49m40.84s & 1.61 $ \times $ 0.71$^*$ & 119.90 & 0.010 & 1.36 & 19.84 & 0.11 & 1.52e+22 \\
N$-$45$^*$ & 18h18m11.49s & -16d49m38.29s & 2.43 $ \times $ 0.86$^*$ & -33.72 & 0.014 & 4.78 & 18.01 & 0.45 & 3.35e+22 \\
N$-$46$^*$ & 18h18m11.35s & -16d49m38.64s & 1.1 $ \times $ 0.55$^*$ & -42.48 & 0.007 & 2.64 & 19.84 & 0.22 & 5.64e+22 \\
N$-$47$^*$ & 18h18m13.68s & -16d49m37.91s & 1.41 $ \times $ 0.9$^*$ & 14.22 & 0.011 & 1.73 & 17.80 & 0.17 & 2.03e+22 \\
N$-$49 & 18h18m12.41s & -16d49m28.08s & 7.37 $ \times $ 6.02 & 111.44 & 0.064 & 790.58 & 21.14 & 60.76 & 2.11e+23 \\
N$-$50 & 18h18m12.33s & -16d49m27.75s & 5.98 $ \times $ 3.24 & 12.07 & 0.042 & 677.29 & 22.33 & 48.55 & 3.85e+23 \\
N$-$51$^*$ & 18h18m11.62s & -16d49m36.52s & 1.02 $ \times $ 0.56$^*$ & -39.81 & 0.007 & 0.79 & 18.24 & 0.07 & 1.99e+22 \\
N$-$52$^*$ & 18h18m13.70s & -16d49m35.33s & 1.05 $ \times $ 0.55$^*$ & 29.18 & 0.007 & 0.60 & 15.50 & 0.07 & 1.87e+22 \\
N$-$53 & 18h18m12.87s & -16d49m30.03s & 5.65 $ \times $ 3.2 & 18.73 & 0.041 & 104.57 & 18.88 & 9.30 & 7.92e+22 \\
N$-$54$^*$ & 18h18m11.45s & -16d49m34.58s & 0.96 $ \times $ 0.47$^*$ & -37.15 & 0.006 & 0.90 & 17.99 & 0.08 & 2.88e+22 \\
N$-$55$^*$ & 18h18m12.92s & -16d49m33.74s & 1.38 $ \times $ 0.9$^*$ & 52.02 & 0.011 & 4.71 & 17.63 & 0.46 & 5.68e+22 \\
N$-$56$^*$ & 18h18m11.55s & -16d49m33.75s & 1.19 $ \times $ 0.55$^*$ & 131.19 & 0.008 & 1.04 & 17.99 & 0.10 & 2.32e+22 \\
N$-$57 & 18h18m12.86s & -16d49m29.61s & 5.29 $ \times $ 2.65 & 30.14 & 0.036 & 86.00 & 19.02 & 7.58 & 8.34e+22 \\
N$-$58$^*$ & 18h18m12.73s & -16d49m32.47s & 1.43 $ \times $ 1.25$^*$ & -41.72 & 0.013 & 15.05 & 19.32 & 1.30 & 1.12e+23 \\
N$-$59 & 18h18m12.32s & -16d49m27.89s & 4.34 $ \times $ 2.47 & -6.19 & 0.031 & 577.63 & 22.80 & 40.34 & 5.81e+23 \\
N$-$60 & 18h18m12.90s & -16d49m28.70s & 3.03 $ \times $ 2.52 & -15.93 & 0.027 & 55.78 & 19.08 & 4.89 & 9.87e+22 \\
N$-$61$^*$ & 18h18m13.03s & -16d49m30.06s & 1.38 $ \times $ 0.77$^*$ & -41.82 & 0.010 & 7.60 & 18.59 & 0.69 & 1.00e+23 \\
N$-$62$^*$ & 18h18m12.86s & -16d49m28.42s & 2.97 $ \times $ 0.28 & 22.33 & 0.009 & 37.67 & 19.18 & 3.28 & 6.14e+23 \\
N$-$63$^*$ & 18h18m12.33s & -16d49m27.93s & 1.51 $ \times $ 0.94 & 21.73 & 0.011 & 288.19 & 22.54 & 20.42 & 2.23e+24 \\
N$-$64 & 18h18m13.47s & -16d49m22.27s & 5.33 $ \times $ 4.71 & 133.24 & 0.048 & 61.73 & 15.93 & 6.90 & 4.23e+22 \\
N$-$65 & 18h18m13.45s & -16d49m22.32s & 4.85 $ \times $ 4.08 & 127.69 & 0.043 & 50.84 & 15.88 & 5.71 & 4.44e+22 \\
N$-$66$^*$ & 18h18m13.99s & -16d49m25.25s & 2.07 $ \times $ 0.86$^*$ & -34.50 & 0.013 & 4.40 & 15.24 & 0.52 & 4.51e+22 \\
N$-$67$^*$ & 18h18m12.29s & -16d49m25.33s & 1.02 $ \times $ 0.58$^*$ & 118.48 & 0.007 & 23.90 & 23.27 & 1.63 & 4.21e+23 \\
N$-$68$^*$ & 18h18m13.37s & -16d49m23.33s & 2.51 $ \times $ 1.03 & 82.56 & 0.015 & 13.08 & 16.20 & 1.43 & 8.45e+22 \\
N$-$69$^*$ & 18h18m11.50s & -16d49m23.33s & 1.18 $ \times $ 0.69$^*$ & -22.91 & 0.009 & 1.21 & 22.35 & 0.09 & 1.64e+22 \\
N$-$70$^*$ & 18h18m12.54s & -16d49m22.31s & 1.73 $ \times $ 1.09$^*$ & 88.83 & 0.013 & 23.54 & 22.46 & 1.68 & 1.37e+23 \\
N$-$71$^*$ & 18h18m13.54s & -16d49m21.99s & 1.06 $ \times $ 0.7$^*$ & -34.98 & 0.008 & 2.73 & 15.83 & 0.31 & 6.40e+22 \\
N$-$72 & 18h18m13.48s & -16d49m20.75s & 3.15 $ \times $ 1.05$^*$ & -33.38 & 0.017 & 8.00 & 15.73 & 0.91 & 4.23e+22 \\
N$-$73$^*$ & 18h18m11.82s & -16d49m21.18s & 1.2 $ \times $ 0.52$^*$ & 103.12 & 0.008 & 0.84 & 22.41 & 0.06 & 1.46e+22 \\
N$-$74$^*$ & 18h18m13.44s & -16d49m19.94s & 1.51 $ \times $ 0.97$^*$ & 3.15 & 0.012 & 4.56 & 15.62 & 0.52 & 5.51e+22 \\
N$-$77$^*$ & 18h18m12.17s & -16d49m12.09s & 3.29 $ \times $ 2.82 & -26.28 & 0.029 & 30.75 & 24.51 & 1.96 & 3.26e+22 \\
N$-$78$^*$ & 18h18m13.09s & -16d49m13.36s & 2.47 $ \times $ 1.38$^*$ & 35.32 & 0.018 & 6.24 & 16.82 & 0.65 & 2.92e+22 \\
N$-$79 & 18h18m13.18s & -16d49m13.12s & 4.54 $ \times $ 1.72 & 78.50 & 0.027 & 12.18 & 16.28 & 1.32 & 2.60e+22 \\
N$-$80 & 18h18m12.18s & -16d49m11.87s & 3.85 $ \times $ 2.87 & 0.64 & 0.032 & 34.23 & 24.51 & 2.18 & 3.04e+22 \\
N$-$81$^*$ & 18h18m13.31s & -16d49m12.09s & 1.39 $ \times $ 0.71$^*$ & 93.25 & 0.010 & 2.81 & 15.77 & 0.32 & 4.96e+22 \\
N$-$82$^*$ & 18h18m11.51s & -16d49m11.56s & 2.17 $ \times $ 0.75$^*$ & 106.28 & 0.012 & 4.11 & 22.07 & 0.30 & 2.85e+22 \\
N$-$84 & 18h18m12.89s & -16d49m07.90s & 3.75 $ \times $ 1.64 & -31.24 & 0.024 & 22.08 & 19.21 & 1.92 & 4.81e+22 \\
N$-$85$^*$ & 18h18m12.92s & -16d49m10.67s & 1.42 $ \times $ 0.36$^*$ & 116.53 & 0.007 & 1.12 & 19.43 & 0.10 & 2.86e+22 \\
N$-$86 & 18h18m12.88s & -16d49m07.50s & 3.1 $ \times $ 1.37 & -42.59 & 0.020 & 18.74 & 19.21 & 1.63 & 5.91e+22 \\
N$-$89$^*$ & 18h18m12.93s & -16d49m08.93s & 1.54 $ \times $ 0.7$^*$ & 91.92 & 0.010 & 2.95 & 18.99 & 0.26 & 3.70e+22 \\
N$-$93$^*$ & 18h18m12.86s & -16d49m06.76s & 2.04 $ \times $ 0.66 & 113.33 & 0.011 & 10.05 & 18.99 & 0.89 & 1.01e+23 \\
\hline
\hline
S$-$2$^*$ & 18h18m13.01s & -16d57m43.32s & 0.84 $ \times $ 0.75$^*$ & 72.09 & 0.008 & 1.76 & 18.51 & 0.16 & 3.92e+22 \\
S$-$3 & 18h18m13.68s & -16d57m40.72s & 4.52 $ \times $ 0.91$^*$ & 53.95 & 0.020 & 8.92 & 18.51 & 0.81 & 3.04e+22 \\
S$-$4$^*$ & 18h18m13.14s & -16d57m42.29s & 1.53 $ \times $ 0.81$^*$ & -42.09 & 0.011 & 2.02 & 18.51 & 0.18 & 2.29e+22 \\
S$-$5$^*$ & 18h18m12.93s & -16d57m42.05s & 1.47 $ \times $ 0.74$^*$ & 67.45 & 0.010 & 1.80 & 18.51 & 0.16 & 2.31e+22 \\
S$-$6$^*$ & 18h18m13.58s & -16d57m41.90s & 1.0 $ \times $ 0.46$^*$ & 89.97 & 0.007 & 2.03 & 18.51 & 0.19 & 6.21e+22 \\
S$-$7 & 18h18m13.63s & -16d57m41.51s & 2.18 $ \times $ 0.63$^*$ & 62.75 & 0.011 & 4.08 & 18.51 & 0.37 & 4.17e+22 \\
S$-$8$^*$ & 18h18m13.68s & -16d57m41.08s & 0.84 $ \times $ 0.65$^*$ & 60.11 & 0.007 & 1.86 & 18.51 & 0.17 & 4.77e+22 \\
S$-$9$^*$ & 18h18m13.76s & -16d57m39.88s & 0.81 $ \times $ 0.69$^*$ & 104.13 & 0.007 & 2.33 & 18.51 & 0.21 & 5.89e+22 \\
S$-$10$^*$ & 18h18m14.10s & -16d57m38.56s & 0.93 $ \times $ 0.53$^*$ & -43.43 & 0.007 & 2.25 & 18.51 & 0.21 & 6.44e+22 \\
S$-$11$^*$ & 18h18m13.63s & -16d57m38.43s & 1.79 $ \times $ 0.62$^*$ & 63.35 & 0.010 & 2.29 & 18.51 & 0.21 & 2.94e+22 \\
S$-$12 & 18h18m14.03s & -16d57m33.95s & 7.66 $ \times $ 4.95 & 73.00 & 0.059 & 59.08 & 19.90 & 4.91 & 1.99e+22 \\
S$-$13 & 18h18m14.03s & -16d57m33.85s & 7.72 $ \times $ 4.74 & 73.30 & 0.058 & 55.83 & 19.90 & 4.64 & 1.95e+22 \\
S$-$14 & 18h18m14.05s & -16d57m33.62s & 7.1 $ \times $ 4.54 & 82.77 & 0.054 & 52.60 & 19.74 & 4.42 & 2.11e+22 \\
S$-$15 & 18h18m14.22s & -16d57m33.74s & 5.22 $ \times $ 2.04 & 31.04 & 0.031 & 34.78 & 18.79 & 3.11 & 4.51e+22 \\
S$-$16$^*$ & 18h18m14.19s & -16d57m37.40s & 1.23 $ \times $ 0.85$^*$ & 70.58 & 0.010 & 3.62 & 18.51 & 0.33 & 4.85e+22 \\
S$-$17 & 18h18m14.23s & -16d57m33.44s & 4.8 $ \times $ 1.72 & 37.57 & 0.028 & 28.93 & 18.77 & 2.59 & 4.84e+22 \\
S$-$18 & 18h18m13.85s & -16d57m33.33s & 4.22 $ \times $ 3.04 & 21.28 & 0.034 & 14.95 & 21.06 & 1.15 & 1.39e+22 \\
S$-$19 & 18h18m13.89s & -16d57m33.01s & 4.43 $ \times $ 1.97 & -3.86 & 0.028 & 10.31 & 21.10 & 0.79 & 1.40e+22 \\
S$-$20 & 18h18m13.88s & -16d57m33.11s & 4.12 $ \times $ 2.05 & -2.07 & 0.028 & 12.40 & 21.06 & 0.96 & 1.75e+22 \\
S$-$21$^*$ & 18h18m13.86s & -16d57m35.65s & 2.18 $ \times $ 0.64$^*$ & 32.33 & 0.011 & 2.69 & 20.19 & 0.22 & 2.43e+22 \\
S$-$22 & 18h18m12.97s & -16d57m20.04s & 25.64 $ \times $ 9.7 & 70.03 & 0.151 & 784.09 & 18.51 & 71.59 & 4.43e+22 \\
S$-$23 & 18h18m12.98s & -16d57m19.99s & 26.03 $ \times $ 9.75 & 70.39 & 0.153 & 800.26 & 18.33 & 74.01 & 4.49e+22 \\
S$-$25$^*$ & 18h18m14.15s & -16d57m34.94s & 2.16 $ \times $ 1.18$^*$ & 51.69 & 0.015 & 5.60 & 19.53 & 0.48 & 2.87e+22 \\
S$-$26$^*$ & 18h18m12.40s & -16d57m34.72s & 1.59 $ \times $ 1.08$^*$ & 133.87 & 0.013 & 1.50 & 16.23 & 0.16 & 1.46e+22 \\
S$-$27 & 18h18m12.98s & -16d57m20.00s & 25.99 $ \times $ 9.7 & 70.33 & 0.152 & 796.26 & 18.37 & 73.43 & 4.49e+22 \\
S$-$28$^*$ & 18h18m13.64s & -16d57m34.69s & 1.47 $ \times $ 0.8$^*$ & 82.40 & 0.010 & 1.26 & 20.09 & 0.10 & 1.36e+22 \\
S$-$30$^*$ & 18h18m14.00s & -16d57m33.67s & 1.64 $ \times $ 0.66$^*$ & -4.80 & 0.010 & 2.06 & 21.08 & 0.16 & 2.26e+22 \\
S$-$31 & 18h18m12.98s & -16d57m20.02s & 25.42 $ \times $ 9.55 & 69.93 & 0.150 & 773.85 & 18.60 & 70.23 & 4.45e+22 \\
S$-$32$^*$ & 18h18m14.28s & -16d57m32.90s & 1.56 $ \times $ 0.93 & 130.22 & 0.012 & 9.10 & 18.26 & 0.85 & 8.99e+22 \\
S$-$33$^*$ & 18h18m13.82s & -16d57m33.61s & 1.3 $ \times $ 0.77$^*$ & 70.62 & 0.010 & 1.55 & 20.90 & 0.12 & 1.86e+22 \\
S$-$34 & 18h18m13.06s & -16d57m20.04s & 23.29 $ \times $ 8.24 & 63.55 & 0.133 & 719.17 & 18.92 & 63.80 & 5.12e+22 \\
S$-$35 & 18h18m13.15s & -16d57m19.56s & 19.34 $ \times $ 8.18 & 59.68 & 0.121 & 674.58 & 18.94 & 59.76 & 5.82e+22 \\
S$-$36 & 18h18m13.15s & -16d57m19.59s & 19.22 $ \times $ 8.11 & 59.86 & 0.120 & 667.87 & 18.99 & 58.99 & 5.83e+22 \\
S$-$37$^*$ & 18h18m12.45s & -16d57m32.74s & 1.01 $ \times $ 0.57$^*$ & -43.99 & 0.007 & 0.32 & 17.75 & 0.03 & 8.41e+21 \\
S$-$38 & 18h18m13.15s & -16d57m19.58s & 19.23 $ \times $ 8.12 & 59.82 & 0.120 & 668.64 & 18.99 & 59.06 & 5.83e+22 \\
S$-$39 & 18h18m13.90s & -16d57m31.98s & 3.09 $ \times $ 0.67 & -42.60 & 0.014 & 5.79 & 21.41 & 0.44 & 3.24e+22 \\
S$-$40$^*$ & 18h18m13.86s & -16d57m31.38s & 1.55 $ \times $ 1.01$^*$ & 89.78 & 0.012 & 3.63 & 21.57 & 0.27 & 2.67e+22 \\
S$-$41$^*$ & 18h18m13.57s & -16d57m31.52s & 1.8 $ \times $ 0.68$^*$ & 81.01 & 0.011 & 1.10 & 20.81 & 0.09 & 1.09e+22 \\
S$-$42 & 18h18m11.44s & -16d57m28.74s & 4.97 $ \times $ 1.5 & 108.34 & 0.026 & 41.08 & 17.42 & 4.07 & 8.37e+22 \\
S$-$43$^*$ & 18h18m12.14s & -16d57m29.91s & 2.03 $ \times $ 0.96$^*$ & 60.59 & 0.013 & 2.06 & 18.84 & 0.18 & 1.45e+22 \\
S$-$44 & 18h18m13.15s & -16d57m19.70s & 18.65 $ \times $ 7.9 & 61.29 & 0.117 & 645.79 & 19.27 & 55.93 & 5.84e+22 \\
S$-$45 & 18h18m12.98s & -16d57m20.01s & 25.94 $ \times $ 9.68 & 70.27 & 0.152 & 793.91 & 18.37 & 73.21 & 4.49e+22 \\
S$-$46$^*$ & 18h18m11.57s & -16d57m29.77s & 1.36 $ \times $ 0.78$^*$ & 106.66 & 0.010 & 5.08 & 18.51 & 0.46 & 6.70e+22 \\
S$-$47 & 18h18m12.98s & -16d57m20.02s & 25.46 $ \times $ 9.55 & 69.98 & 0.150 & 774.60 & 18.60 & 70.30 & 4.45e+22 \\
S$-$48$^*$ & 18h18m11.35s & -16d57m28.31s & 1.66 $ \times $ 1.05$^*$ & 132.77 & 0.013 & 18.34 & 18.51 & 1.67 & 1.48e+23 \\
S$-$49 & 18h18m13.15s & -16d57m19.67s & 18.52 $ \times $ 7.9 & 61.42 & 0.116 & 642.37 & 19.28 & 55.59 & 5.85e+22 \\
S$-$50 & 18h18m12.91s & -16d57m22.27s & 11.45 $ \times $ 5.33 & 93.07 & 0.075 & 417.16 & 20.87 & 32.60 & 8.23e+22 \\
S$-$51 & 18h18m12.92s & -16d57m22.27s & 11.15 $ \times $ 5.25 & 92.98 & 0.073 & 404.12 & 21.00 & 31.32 & 8.23e+22 \\
S$-$52 & 18h18m12.46s & -16d57m24.04s & 3.72 $ \times $ 2.62 & -17.70 & 0.030 & 73.49 & 21.38 & 5.57 & 8.78e+22 \\
S$-$53 & 18h18m12.46s & -16d57m25.76s & 2.33 $ \times $ 0.95 & 83.95 & 0.014 & 21.97 & 20.60 & 1.74 & 1.21e+23 \\
S$-$54$^*$ & 18h18m12.43s & -16d57m25.79s & 1.62 $ \times $ 1.44$^*$ & 40.14 & 0.015 & 17.92 & 20.56 & 1.43 & 9.39e+22 \\
S$-$55$^*$ & 18h18m14.72s & -16d57m25.99s & 1.08 $ \times $ 0.68$^*$ & 46.59 & 0.008 & 1.17 & 19.21 & 0.10 & 2.14e+22 \\
S$-$56$^*$ & 18h18m12.62s & -16d57m25.62s & 0.93 $ \times $ 0.53$^*$ & 103.14 & 0.007 & 2.65 & 20.69 & 0.21 & 6.60e+22 \\
S$-$57 & 18h18m13.06s & -16d57m21.85s & 8.88 $ \times $ 2.07 & 116.48 & 0.041 & 276.57 & 20.98 & 21.46 & 1.80e+23 \\
S$-$58 & 18h18m13.26s & -16d57m23.35s & 3.86 $ \times $ 1.05 & 119.85 & 0.019 & 110.62 & 20.95 & 8.60 & 3.28e+23 \\
S$-$59$^*$ & 18h18m13.34s & -16d57m23.97s & 1.52 $ \times $ 0.95$^*$ & 98.28 & 0.012 & 52.65 & 20.79 & 4.13 & 4.41e+23 \\
S$-$60$^*$ & 18h18m11.33s & -16d57m24.02s & 1.26 $ \times $ 0.77$^*$ & 53.66 & 0.009 & 0.73 & 18.51 & 0.07 & 1.05e+22 \\
S$-$61$^*$ & 18h18m12.43s & -16d57m22.71s & 1.61 $ \times $ 0.55 & 108.42 & 0.009 & 33.11 & 21.94 & 2.43 & 4.21e+23 \\
S$-$62$^*$ & 18h18m13.16s & -16d57m22.70s & 1.51 $ \times $ 0.55$^*$ & -28.06 & 0.009 & 12.68 & 20.90 & 0.99 & 1.84e+23 \\
S$-$63 & 18h18m11.35s & -16d57m19.66s & 6.97 $ \times $ 2.79 & 102.31 & 0.042 & 49.01 & 15.00 & 5.95 & 4.72e+22 \\
S$-$64$^*$ & 18h18m12.16s & -16d57m20.92s & 1.91 $ \times $ 0.94$^*$ & 91.10 & 0.013 & 4.14 & 19.35 & 0.36 & 3.08e+22 \\
S$-$65$^*$ & 18h18m12.84s & -16d57m20.20s & 1.73 $ \times $ 0.77 & 106.81 & 0.011 & 80.77 & 21.61 & 6.04 & 6.99e+23 \\
S$-$66$^*$ & 18h18m14.77s & -16d57m20.28s & 1.21 $ \times $ 0.51$^*$ & -37.61 & 0.008 & 0.67 & 15.03 & 0.08 & 2.03e+22 \\
S$-$67$^*$ & 18h18m11.49s & -16d57m20.02s & 1.52 $ \times $ 0.59$^*$ & 99.97 & 0.009 & 2.73 & 15.56 & 0.32 & 5.43e+22 \\
S$-$68$^*$ & 18h18m11.25s & -16d57m19.20s & 2.11 $ \times $ 1.03$^*$ & 85.35 & 0.014 & 8.99 & 16.81 & 0.93 & 6.61e+22 \\
S$-$69 & 18h18m13.73s & -16d57m13.10s & 8.86 $ \times $ 5.72 & 94.21 & 0.068 & 185.50 & 16.91 & 19.12 & 5.81e+22 \\
S$-$70 & 18h18m13.76s & -16d57m12.79s & 8.59 $ \times $ 4.73 & 102.58 & 0.061 & 158.26 & 16.72 & 16.56 & 6.28e+22 \\
S$-$71$^*$ & 18h18m13.40s & -16d57m17.68s & 1.18 $ \times $ 0.41 & 129.66 & 0.007 & 6.73 & 19.00 & 0.59 & 1.89e+23 \\
S$-$72 & 18h18m13.91s & -16d57m13.17s & 5.04 $ \times $ 2.77 & 2.62 & 0.036 & 105.76 & 16.84 & 10.96 & 1.21e+23 \\
S$-$73 & 18h18m13.77s & -16d57m12.71s & 8.44 $ \times $ 4.45 & 104.69 & 0.059 & 143.17 & 16.63 & 15.10 & 6.19e+22 \\
S$-$74$^*$ & 18h18m13.95s & -16d57m17.06s & 1.06 $ \times $ 0.96$^*$ & 49.97 & 0.010 & 5.40 & 15.57 & 0.62 & 9.40e+22 \\
S$-$75$^*$ & 18h18m14.96s & -16d57m16.56s & 1.72 $ \times $ 0.62$^*$ & 109.14 & 0.010 & 1.40 & 14.69 & 0.18 & 2.54e+22 \\
S$-$76$^*$ & 18h18m14.54s & -16d57m14.80s & 2.6 $ \times $ 0.79 & 105.96 & 0.014 & 9.73 & 15.17 & 1.16 & 8.77e+22 \\
S$-$77$^*$ & 18h18m13.92s & -16d57m12.34s & 3.06 $ \times $ 1.51 & 1.88 & 0.021 & 69.08 & 16.81 & 7.18 & 2.39e+23 \\
S$-$78$^*$ & 18h18m13.38s & -16d57m14.69s & 1.03 $ \times $ 0.83$^*$ & -40.87 & 0.009 & 3.09 & 18.67 & 0.28 & 5.02e+22 \\
S$-$79$^*$ & 18h18m12.63s & -16d57m13.07s & 2.69 $ \times $ 0.84$^*$ & 52.38 & 0.014 & 1.31 & 18.91 & 0.12 & 7.94e+21 \\
S$-$80$^*$ & 18h18m13.38s & -16d57m11.45s & 2.68 $ \times $ 1.65 & 60.53 & 0.020 & 35.50 & 15.99 & 3.95 & 1.37e+23 \\
S$-$81$^*$ & 18h18m15.29s & -16d57m09.81s & 1.37 $ \times $ 0.48$^*$ & -20.85 & 0.008 & 1.92 & 18.51 & 0.17 & 4.09e+22 \\
S$-$82$^*$ & 18h18m12.19s & -16d57m09.95s & 1.24 $ \times $ 0.72$^*$ & -13.28 & 0.009 & 0.82 & 18.51 & 0.08 & 1.29e+22 \\
S$-$83$^*$ & 18h18m12.41s & -16d57m09.16s & 0.99 $ \times $ 0.71$^*$ & -1.87 & 0.008 & 0.50 & 18.51 & 0.05 & 9.91e+21 \\
S$-$84 & 18h18m12.43s & -16d57m05.69s & 5.88 $ \times $ 2.43 & 44.14 & 0.036 & 15.58 & 18.51 & 1.42 & 1.53e+22 \\
S$-$85 & 18h18m12.44s & -16d57m05.96s & 6.69 $ \times $ 2.82 & 48.67 & 0.042 & 18.21 & 18.51 & 1.66 & 1.36e+22 \\
S$-$86 & 18h18m12.46s & -16d57m05.69s & 6.0 $ \times $ 2.87 & 51.21 & 0.040 & 17.19 & 18.51 & 1.57 & 1.40e+22 \\
S$-$87 & 18h18m12.44s & -16d57m05.35s & 5.82 $ \times $ 1.87 & 49.84 & 0.032 & 13.19 & 18.51 & 1.20 & 1.70e+22 \\
S$-$88$^*$ & 18h18m13.70s & -16d57m07.97s & 1.05 $ \times $ 0.52$^*$ & -10.33 & 0.007 & 0.47 & 16.27 & 0.05 & 1.42e+22 \\
S$-$89$^*$ & 18h18m13.09s & -16d57m07.94s & 1.05 $ \times $ 0.55$^*$ & 79.01 & 0.007 & 0.54 & 18.51 & 0.05 & 1.32e+22 \\
S$-$90 & 18h18m12.31s & -16d57m06.54s & 2.13 $ \times $ 0.22 & 31.94 & 0.007 & 4.86 & 18.51 & 0.44 & 1.48e+23 \\
S$-$91 & 18h18m13.89s & -16d57m05.42s & 2.99 $ \times $ 0.42 & -7.14 & 0.011 & 7.85 & 18.08 & 0.74 & 9.02e+22 \\
S$-$92$^*$ & 18h18m12.27s & -16d57m07.43s & 0.82 $ \times $ 0.62$^*$ & -3.41 & 0.007 & 0.99 & 18.51 & 0.09 & 2.74e+22 \\
S$-$93$^*$ & 18h18m13.88s & -16d57m06.42s & 1.42 $ \times $ 0.79$^*$ & -39.96 & 0.010 & 2.23 & 18.26 & 0.21 & 2.85e+22 \\
S$-$94$^*$ & 18h18m12.31s & -16d57m06.21s & 1.39 $ \times $ 0.41$^*$ & 106.46 & 0.007 & 0.92 & 18.51 & 0.08 & 2.26e+22 \\
S$-$95$^*$ & 18h18m12.74s & -16d57m05.37s & 1.22 $ \times $ 0.73$^*$ & 56.95 & 0.009 & 0.86 & 18.51 & 0.08 & 1.36e+22 \\
S$-$96 & 18h18m12.56s & -16d57m03.67s & 2.29 $ \times $ 1.1 & 70.44 & 0.015 & 5.69 & 18.51 & 0.52 & 3.17e+22 \\
S$-$97$^*$ & 18h18m13.89s & -16d57m04.18s & 1.3 $ \times $ 0.85$^*$ & -27.18 & 0.010 & 2.62 & 17.01 & 0.27 & 3.72e+22 \\
S$-$98$^*$ & 18h18m12.60s & -16d57m03.40s & 1.69 $ \times $ 0.74$^*$ & -1.04 & 0.011 & 2.25 & 18.51 & 0.21 & 2.53e+22 \\
S$-$99$^*$ & 18h18m12.47s & -16d57m03.90s & 1.19 $ \times $ 0.61$^*$ & 29.76 & 0.008 & 1.36 & 18.51 & 0.12 & 2.63e+22 \\
S$-$100$^*$ & 18h18m12.69s & -16d57m00.14s & 1.37 $ \times $ 0.45$^*$ & 63.44 & 0.008 & 0.97 & 18.51 & 0.09 & 2.21e+22 \\
S$-$101 & 18h18m13.72s & -16d56m57.24s & 3.76 $ \times $ 2.66 & 127.26 & 0.030 & 5.83 & 18.51 & 0.53 & 8.21e+21 \\
S$-$102$^*$ & 18h18m13.79s & -16d56m58.26s & 2.17 $ \times $ 0.98$^*$ & -1.61 & 0.014 & 2.21 & 18.51 & 0.20 & 1.46e+22 \\
S$-$104$^*$ & 18h18m13.62s & -16d56m56.78s & 1.21 $ \times $ 0.9$^*$ & -20.59 & 0.010 & 1.13 & 18.51 & 0.10 & 1.46e+22 \\
\hline
\end{longtable*}
%\end{center}

\begin{center}
\begin{longtable*}[ht!]{c|c|c|c|c}
\caption{Dynamical parameters of core structures} 
\label{tab:continuum_dyn_para} \\
\hline \hline \multicolumn{1}{c|}{id} & \multicolumn{1}{c|}{$\Delta v$} & \multicolumn{1}{c|}{$M_{vir}$} & \multicolumn{1}{c|}{$\alpha$} & \multicolumn{1}{c}{$\mathcal{M}$} \\
 & (km~s$^{-1}$) & ($M_\odot$) & & \\
\endfirsthead

\multicolumn{5}{c}% \\
{{ \tablename\ \thetable{} -- Dynamical parameters of core structures (continued)}} \\
\hline \hline \hline \multicolumn{1}{c|}{id} & \multicolumn{1}{c|}{$\Delta v$} & \multicolumn{1}{c|}{$M_{vir}$} & \multicolumn{1}{c|}{$\alpha$} & \multicolumn{1}{c}{$\mathcal{M}$} \\
 & (km~s$^{-1}$) & ($M_\odot$) & & \\
\hline 
\endhead
\hline 
\multicolumn{5}{r}{$\Delta v$ is the N$_2$D$^+$ linewidth of the identified structure. $\Delta v$ with * cannot be deconvolved with the channel width.} \\
\multicolumn{5}{r}{$M_{vir}$ is virial mass, $\alpha$ represents virial parameter, and $\mathcal{M}$ represents Mach number.} \\
\multicolumn{5}{r}{... in $\mathcal{M}$ column means that Mach number is not available for structures with $\sigma_V<\sigma_{th}$.} \\
\multicolumn{5}{r}{{Continued on next page}} \\
\endfoot
\hline 
\endlastfoot

\hline
N.4$-$0 & 1.22 & 19.21 & 1.50 & 3.20 \\
N.4$-$1 & 1.22 & 18.12 & 1.48 & 3.16 \\
N.4$-$2 & 1.24 & 14.04 & 1.56 & 3.10 \\
N.4$-$3 & 0.86 & 1.59 & 7.84 & 2.33 \\
N.4$-$5 & 1.23 & 13.26 & 1.55 & 3.07 \\
N.4$-$6 & 0.98 & 1.90 & 13.27 & 2.70 \\
N.4$-$7 & 1.20 & 11.72 & 1.46 & 2.99 \\
N.4$-$8 & 1.11 & 7.33 & 1.12 & 2.65 \\
N.4$-$9 & 1.14 & 3.00 & 2.39 & 3.02 \\
N.4$-$11 & 1.15 & 9.01 & 1.24 & 2.82 \\
N.4$-$12 & 1.10 & 4.09 & 0.87 & 2.56 \\
N.4$-$14 & 1.41 & 2.94 & 22.76 & 4.20 \\
N.4$-$16 & 0.82 & 1.78 & 25.61 & 1.82 \\
\hline
N.4$-$4 & 0.27 $\ $  0.74 & 0.13 $\ $  0.97 &  5.73 $\ $  16.12 & 0.61 $\ $  2.09 \\
N.4$-$13 & 0.56 $\ $  0.71 &  0.67 $\ $  1.08 &  6.55 $\ $  8.26 & 1.52 $\ $  1.96 \\
N.4$-$17 & 0.44 $\ $  0.87 &  0.32 $\ $  1.26 & 17.92 $\ $  23.82 & 1.04 $\ $  2.21 \\
\hline
\hline
N.5$-$0 & 0.88 & 18.48 & 2.36 & 2.82 \\
N.5$-$1 & 0.59 & 1.79 & 1.33 &1.90 \\
N.5$-$2 & 0.63 & 0.71 & 4.41 & 2.00 \\
N.5$-$3 & 0.62 & 1.44 & 3.80 & 1.99 \\
N.5$-$4 & 0.87 & 10.08 & 1.90 & 2.73 \\
N.5$-$5 & 0.57 & 0.69 & 0.62 & 1.81 \\
N.5$-$6 & 0.78 & 5.93 & 1.71 & 2.44 \\
N.5$-$7 & 0.17 & 0.09 & 0.43 & 0.28 \\
N.5$-$8 & 0.88 & 6.55 & 2.14 & 2.77 \\
N.5$-$9 & 0.86 & 5.12 & 1.84 & 2.76 \\
N.5$-$10 & 0.40 & 0.24 & 0.67 & 1.20 \\
\hline
\hline
N.7$-$1 & 0.94 & 1.56 & 6.81 & 3.03 \\
N.7$-$2 & 0.82 & 2.22 & 4.65 & 2.65 \\
\hline
N.7$-$0 & 0.89 $\ $ 0.32 &  10.74 $\ $ 1.42 &  2.86 $\ $ 0.37 & 2.88 $\ $ 0.94 \\
N.7$-$3 & 0.79 $\ $ 0.25 &  6.48 $\ $ 0.63 &  2.52 $\ $  0.25  & 2.53 $\ $  0.64 \\
N.7$-$4 & 0.19$^*$  $\ $   0.46 & 0.18 $\ $  1.04 & 0.13 $\ $  0.68 & ... $\ $  1.42 \\
N.7$-$5 & 0.73   $\ $  0.19 &  1.23 $\ $  0.09 &  1.81 $\ $  0.26 & 2.34 $\ $  0.40 \\
N.7$-$6 & 0.66  $\ $   0.26 &  2.00  $\ $  0.31 & 3.70 $\ $  1.62 &  2.10 $\ $  0.70 \\
N.7$-$7 & 0.56   $\ $  0.20$^*$ &  0.62  $\ $  0.08  & 10.91 $\ $  0.80 & 1.76  $\ $  ... \\
\hline
\hline
N$-$3 & 0.58 & 1.00 & 0.94 & 1.78 \\
N$-$5 & 0.27 & 0.05 & 0.06 & 0.67 \\
N$-$6 & 0.58 & 0.80 & 1.56 & 1.70 \\
N$-$7$^*$ & 0.18 & 0.06 & 0.31 & ... \\
N$-$14 & 0.32 & 0.26 & 1.56 & 0.86 \\
N$-$15 & 1.37 & 20.06 & 1.71 & 4.26 \\
N$-$18 & 0.89 & 2.99 & 1.36 & 2.70 \\
N$-$24 & 0.59 & 1.24 & 1.76 & 1.60 \\
N$-$26 & 1.01 & 2.45 & 5.98 & 2.80 \\
N$-$30 & 1.17 & 8.79 & 1.03 & 3.30 \\
N$-$31 & 0.32 & 0.18 & 1.06 & 0.82 \\
N$-$32 & 0.61 & 1.77 & 2.54 & 1.59 \\
N$-$35 & 1.25 & 3.75 & 14.66 & 3.58 \\
N$-$36 & 1.02 & 3.45 & 0.55 & 2.85 \\
N$-$41 & 0.67 & 1.11 & 2.20 & 1.81 \\
N$-$47 & 0.59 & 0.78 & 4.67 & 1.66 \\
N$-$50 & 1.39 & 17.12 & 0.35 & 3.63 \\
N$-$52 & 1.13 & 1.93 & 27.69 & 3.53 \\
N$-$58 & 0.83 & 1.86 & 1.43 & 2.31 \\
N$-$59 & 1.26 & 10.38 & 0.26 & 3.24 \\
N$-$61 & 0.60 & 0.75 & 1.09 & 1.67 \\
N$-$63 & 1.16 & 3.24 & 0.16 & 3.02 \\
N$-$70 & 0.67 & 1.23 & 0.74 & 1.69 \\
N$-$71 & 0.83 & 1.21 & 3.93 & 2.57 \\
N$-$72 & 1.00 & 3.64 & 3.99 & 3.09 \\
N$-$74 & 0.97 & 2.31 & 4.40 & 3.03 \\
N$-$77 & 1.48 & 13.38 & 6.82 & 3.69 \\
N$-$78 & 0.66 & 1.61 & 2.48 & 1.94 \\
N$-$80 & 1.72 & 19.85 & 9.09 & 4.31 \\
N$-$81 & 0.63 & 0.81 & 2.54 & 1.93 \\
\hline
N$-$1 & 1.06  $\ $  1.29 & 37.33  $\ $  55.10 & 0.74  $\ $  0.89 & 3.03  $\ $  3.69 \\
N$-$2 & 0.69  $\ $  0.72 & 5.20  $\ $  5.58 & 1.84  $\ $  1.47 & 2.21  $\ $  2.29 \\
N$-$4 & 0.26  $\ $  0.90 & 0.38  $\ $  4.47 & 0.34  $\ $  1.43 & 0.71  $\ $  2.89 \\
N$-$8 & 0.70  $\ $  0.72 & 5.33  $\ $  5.59 & 1.84  $\ $  1.45 & 2.23  $\ $  2.28 \\
N$-$9 & 1.09  $\ $  1.16 & 35.23  $\ $  40.07 & 0.69  $\ $  0.77 & 3.07  $\ $  3.28 \\
N$-$10 & 1.09  $\ $  1.17 & 34.95  $\ $  39.86 & 0.70 $\ $  0.79 & 3.07 $\ $  3.28 \\
N$-$12 & 1.09  $\ $  1.18 & 34.77  $\ $  40.46 & 0.71 $\ $  0.82 & 3.06 $\ $  3.31 \\
N$-$13 & 0.76  $\ $  0.52 & 6.41  $\ $  2.97 & 0.89 $\ $  0.59 & 2.32 $\ $  1.54 \\
N$-$16 & 0.63  $\ $  0.42 & 2.09  $\ $  0.92 & 0.65 $\ $  0.43 & 1.96 $\ $  1.25 \\
N$-$20 & 1.11  $\ $  1.12 & 26.78  $\ $  27.39 & 0.59 $\ $ 0.72 & 3.02 $\ $ 3.05 \\
N$-$21 & 1.10  $\ $  1.12 & 26.34 $\ $  27.05 & 0.59 $\ $  0.71 & 3.01 $\ $  3.05 \\
N$-$23 & 1.10  $\ $  1.17 & 23.63 $\ $  26.77 & 0.55 $\ $  0.84 & 2.98 $\ $  3.18 \\
N$-$28 & 1.10  $\ $  1.21 & 22.76 $\ $  27.50 & 0.53 $\ $  0.90 & 2.97 $\ $  3.28 \\
N$-$49 & 1.12  $\ $  1.18 & 16.79 $\ $  18.70 & 0.36 $\ $  1.36 & 2.99 $\ $  3.16 \\
N$-$53 & 0.79  $\ $  0.76 & 5.29 $\ $  4.94 & 1.08 $\ $  1.12 & 2.20 $\ $  2.12 \\
N$-$55 & 0.62  $\ $  0.74 & 0.87 $\ $  1.23 & 7.21 $\ $  3.64 & 1.78 $\ $  2.15 \\
N$-$57 & 0.81 $\ $  0.81 & 5.00 $\ $  4.92 & 1.09 $\ $  1.65 & 2.28 $\ $  2.26 \\
N$-$60 & 0.63  $\ $  0.74 & 2.22 $\ $  3.08 & 0.94 $\ $  1.22 & 1.74 $\ $  2.07 \\
N$-$62 & 0.65  $\ $  0.72 & 0.78 $\ $  0.94 & 0.40 $\ $  0.70 & 1.80 $\ $  1.98 \\
N$-$64 & 1.20  $\ $  0.60 & 14.57 $\ $  3.69 & 2.63 $\ $  2.71 & 3.72 $\ $  1.83 \\
N$-$65 & 1.18  $\ $  0.62 & 12.46 $\ $  3.50 & 2.75 $\ $  2.97 & 3.66 $\ $  1.89 \\
N$-$67 & 1.09  $\ $  1.21 & 1.83 $\ $  2.28 & 1.72 $\ $  4.07 & 2.76 $\ $  3.09 \\
N$-$68 & 1.33  $\ $  0.56 & 5.73 $\ $  1.04 & 5.83 $\ $  2.33 & 4.08 $\ $  1.68 \\
N$-$79 & 0.68  $\ $  0.73 & 2.60 $\ $  3.02 & 3.01 $\ $  6.56 & 2.04 $\ $  2.21 \\
\hline
\hline
S$-$22 & 2.29 & 167.00 & 2.33 & 6.62 \\
S$-$23 & 2.32 & 173.38 & 2.34 & 6.75 \\
S$-$26 & 0.78 & 1.61 & 9.86 & 2.36 \\
S$-$27 & 2.35 & 177.07 & 2.41 & 6.83 \\
S$-$31 & 2.33 & 169.95 & 2.42 & 6.71 \\
S$-$34 & 2.51 & 176.61 & 2.77 & 7.19 \\
S$-$35 & 2.51 & 159.96 & 2.68 & 7.18 \\
S$-$36 & 2.51 & 158.61 & 2.69 & 7.17 \\
S$-$38 & 2.51 & 158.26 & 2.68 & 7.16 \\
S$-$41 & 0.84 & 1.58 & 18.25 & 2.25 \\
S$-$42 & 0.74 & 3.05 & 0.75 & 2.17 \\
S$-$43 & 0.49 & 0.68 & 3.70 & 1.33 \\
S$-$44 & 2.54 & 158.18 & 2.83 & 7.21 \\
S$-$45 & 2.35 & 176.75 & 2.41 & 6.83 \\
S$-$46 & 0.78 & 1.26 & 2.72 & 2.20 \\
S$-$47 & 2.32 & 169.88 & 2.42 & 6.70 \\
S$-$48 & 0.73 & 1.43 & 0.86 & 2.07 \\
S$-$49 & 2.54 & 157.88 & 2.84 & 7.21 \\
S$-$50 & 1.79 & 50.73 & 1.56 & 4.88 \\
S$-$51 & 1.85 & 53.07 & 1.69 & 5.02 \\
S$-$52 & 1.39 & 12.12 & 2.18 & 3.71 \\
S$-$53 & 0.86 & 2.24 & 1.28 & 2.32 \\
S$-$54 & 0.90 & 2.49 & 1.75 & 2.42 \\
S$-$56 & 0.73 & 0.75 & 3.56 & 1.93 \\
S$-$57 & 1.85 & 29.47 & 1.37 & 5.00 \\
S$-$58 & 1.07 & 4.60 & 0.53 & 2.86 \\
S$-$59 & 0.75 & 1.36 & 0.33 & 1.99 \\
S$-$61 & 1.23 & 2.86 & 1.18 & 3.23 \\
S$-$62 & 0.63 & 0.73 & 0.74 & 1.65 \\
S$-$63 & 1.00 & 8.96 & 1.50 & 3.20 \\
S$-$64 & 0.54 & 0.78 & 2.20 & 1.45 \\
S$-$65 & 1.23 & 3.50 & 0.58 & 3.25 \\
S$-$66 & 0.26 & 0.11 & 1.31 & 0.68 \\
S$-$67 & 0.58 & 0.64 & 2.03 & 1.76 \\
S$-$68 & 1.23 & 4.47 & 4.78 & 3.70 \\
S$-$75 & 0.59 & 0.73 & 4.12 & 1.86 \\
S$-$76 & 0.61 & 1.07 & 0.92 & 1.89 \\
S$-$77 & 1.33 & 7.63 & 1.06 & 4.01 \\
S$-$78 & 1.03 & 1.96 & 7.05 & 2.92 \\
S$-$79 & 0.54 & 0.90 & 7.71 & 1.48 \\
S$-$80 & 1.08 & 4.97 & 1.26 & 3.34 \\
S$-$88 & 0.20$^*$ & 0.06 & 1.14 & ...  \\
S$-$91 & 0.67 & 1.02 & 1.39 & 1.91 \\
S$-$93 & 0.50 & 0.53 & 2.54 & 1.37 \\
\hline
S$-$69 & 0.90 $\ $ 1.00 $\ $  1.19 & 11.63 $\ $  14.49  $\ $  20.32 & 2.83  $\ $  3.10  $\ $  1.97 & 2.69  $\ $  3.01  $\ $  3.58 \\
S$-$70 & 0.90  $\ $  1.20  $\ $  1.09 & 10.32  $\ $  18.36  $\ $  15.13 & 3.80  $\ $  3.26  $\ $  1.84 & 2.69  $\ $  3.61  $\ $  3.27 \\
S$-$71 & 0.18$^*$  $\ $  0.54 & 0.05  $\ $  0.40 & 0.19  $\ $  1.14 & ...  $\ $  1.46 \\
S$-$72 & 2.15  $\ $  0.52 & 34.80  $\ $  2.07 & 3.50 $\ $  2.05 & 6.52 $\ $  1.52 \\
S$-$73 & 0.91  $\ $  1.23  $\ $  1.06 & 10.18  $\ $  18.68  $\ $  13.92 & 4.53  $\ $  3.36  $\ $  1.91 & 2.73  $\ $  3.73  $\ $  3.21 \\
S$-$74 & 1.17  $\ $  1.01 & 2.78  $\ $  2.06 & 7.11  $\ $  8.92 & 3.66  $\ $  3.14 \\
S$-$89 & 0.53  $\ $  0.47 & 0.43  $\ $  0.34 & 18.16  $\ $  13.03 & 1.46  $\ $  1.27\\
\hline
\end{longtable*}
\end{center}

\newpage

\begin{table*}[t!]
\caption{Outflow parameters}
\label{tab:outflow_parameter}
\centering
\begin{tabular}{c|ccccc}
\hline
\hline
Field & Outflow & Mass   & Momentum & Energy & Integrted velocity range \\
      &   &  ($M_\odot$) & ($M_\odot \ km \cdot s^{-1}$) & ($M_\odot \ km^2 \cdot s^{-2}$) & ($km\cdot s^{-1}$) \\
\hline
N.4 field & 1       & 0.0069 & 0.018    & 0.027  & 15.0$-$19.2\\
& 2       & 0.0050 & 0.011    & 0.014  & 15.6$-$18.6\\
& 3       & 0.0036 & 0.0043   & 0.0026 & 18.3$-$19.2\\
& 4       & 0.0020 & 0.0084   & 0.018  & 23.4$-$27.0\\
& 5       & 0.022  & 0.11     & 0.31  & 23.4$-$29.4\\
\hline
N.5 field & 1       & 0.045  & 0.37     & 2.2  & -10.8$-$18.0\\
& 2       & 0.0037 & 0.024    & 0.077  & 11.7$-$15.0\\
& 3       & 0.012  & 0.093    & 0.37   & 7.8$-$14.4\\
& red     & 0.051  & 0.30    & 0.98    & 23.1$-$30.6\\
%4       & 0.012  & 0.080    & 0.34   \\
%5       & 0.021  & 0.12     & 0.36  \\
\hline
N.7 field & 1       & 0.076 & 0.48     & 2.3  & -4.2$-$18.3\\
& 2       & 0.025 & 0.28     & 2.0   & 24.6$-$47.1\\
\hline
Mosaic N field & 1a      & 0.094   & 0.63     & 2.3    & 3.9$-$15.3\\
& 1b      & 0.12    & 0.83     & 3.2    & -9.3$-$16.2\\
& 1c      & 0.34    & 0.39     & 2.9    & -16.2$-$15.3\\
& 1d      & 0.0061  & 0.048    & 0.18   & -17.7$-$14.7\\
& 1e      & 0.017   & 0.15     & 0.83   & 24.6$-$44.4\\
& 1f      & 0.028   & 0.14     & 0.39   & 23.1$-$28.2\\    
& 1g      & 0.0070  & 0.053    & 0.22   & 23.4$-$34.8\\
& 1h      & 0.073   & 0.61     & 3.9    & 34.6$-$57.6\\
& 1i      & 0.021   & 0.14     & 0.62   & 22.8$-$40.5\\
& 2a      & 0.013   & 0.15     & 1.5    & -18.0$-$16.5\\
&        & 0.021   & 0.17     & 0.87   & 22.8$-$43.2\\
& 3a      & 0.0055  & 0.039    & 0.18   & -3.9$-$15.9\\
&        & 0.00071 & 0.0079   & 0.046  & 27.9$-$37.5\\
\hline
Mosaic S field & 1a      & 0.11   & 1.3      & 12     &  -18.0$-$15.9\\
&         & 0.029  & 0.50     & 5.7    & 24.3$-$57.3\\
& 1b      & 0.011  & 0.077    & 0.28   & 25.5$-$30.3\\
&         & 0.0052 & 0.037    & 0.14   & 24.3$-$30.9\\
& 1c      & 0.014  & 0.093    & 0.33   & 24.0$-$30.3\\
& 2a      & 0.0012 & 0.0075   & 0.025  & 9.0$-$16.5\\
&         & 0.0033 & 0.022    & 0.075  & 23.7$-$32.1\\
& 2b      & 0.0034 & 0.021    & 0.069  & 10.5$-$15.9\\
& 3a      & 0.010  & 0.074    & 0.33   & -2.93$-$16.5\\
& 3b      & 0.0013 & 0.0096   & 0.038  & 24.3$-$33.3\\
& 4a      & 0.015  & 0.082    & 0.26   & 5.4$-$16.5\\
& 5a      & 0.0092 & 0.088    & 0.54   & -4.2$-$16.5\\
& 5b      & 0.017  & 0.18     & 1.0    & 27.3$-$36.3\\
& 5c      & 0.0062 & 0.051    & 0.22   & 25.8$-$36.3\\
& 6a      & 0.014  & 0.076    & 0.23   & 9.9$-$16.5\\
& 6b      & 0.0085 & 0.078    & 0.40   & 24.3$-$37.5\\
& 7a      & 0.073  & 0.77     & 5.4    & -18.0$-$16.5\\
& 7b      & 0.035  & 0.45     & 3.6    & 25.5$-$52.5\\
& 8a      & 0.039  & 0.50     & 4.6    & -18.0$-$16.2 \\
\hline
\hline
\end{tabular}
\end{table*}



\end{document}